\documentclass[preprint,12pt]{elsarticle}

\usepackage{amssymb}
\usepackage{nomencl}
\usepackage{gensymb}
\usepackage{graphicx}
\usepackage{multirow}
\usepackage{algpseudocode}
\usepackage[hang,flushmargin]{footmisc} 
\usepackage{subfig}
\usepackage{amsmath}
\usepackage{setspace} 
\usepackage{babel}
\usepackage{blindtext}
\usepackage[linesnumbered,ruled,vlined]{algorithm2e}

\journal{Wind Engineering and Industrial Aerodynamics}

\begin{document}
\emergencystretch 3em

\begin{frontmatter}



\title{A Novel Framework for Optimizing Gurney Flaps using RBF Neural Network and Cuckoo Search Algorithm}

\author[inst1]{Aryan Tyagi\textsuperscript{\textdagger}}
\author[inst1]{Paras Singh\textsuperscript{\textdagger}}
\author[inst2]{Aryaman Rao}
\author[inst1]{\\Gaurav Kumar\textsuperscript{*}}
\author[inst1]{Raj Kumar Singh}


\affiliation[inst1]{organization={Department of Mechanical Engineering, Delhi Technological University},
            city={New Delhi},
            country={India}}
            
\affiliation[inst2]{organization={Department of Electrical Engineering, Delhi Technological University},
            city={New Delhi},
            country={India}}

\def\thefootnote{\textdagger}
\footnotetext{The authors contributed equally and would like to be considered as joint first authors.}
\def\thefootnote{\arabic{footnote}}
\def\thefootnote{*}\footnotetext{Corresponding author;
  E-mail address: gauravkmr716@gmail.com;\\
  Phone: +91 8750700353}\def\thefootnote{\arabic{footnote}}


\begin{abstract}
Enhancing aerodynamic efficiency is vital for optimizing aircraft performance and operational effectiveness. It enables greater speeds and reduced fuel consumption, leading to lower operating costs. Hence, the implementation of Gurney flaps represents a promising avenue for improving airfoil aerodynamics. The optimization of Gurney flaps holds considerable ramifications for improving the lift and stall characteristics of airfoils in aircraft and wind turbine blade designs. The efficacy of implementing Gurney flaps hinges significantly on its design parameters, namely, flap height and mounting angle. This study attempts to optimize these parameters using a design optimization framework, which incorporates training a Radial Basis Function surrogate model based on CFD data from two-dimensional (2D) Reynolds-Averaged Navier-Stokes (RANS) simulations. The Cuckoo Search algorithm is then employed to obtain the optimal design parameters and compared with other competing optimization algorithms. The optimized Gurney flap configuration shows a notable improvement of 10.28\% in $C_l/C_d$, with a flap height of 1.9\%c and a flap mounting angle of $-58\degree$. The study highlights the effectiveness of the proposed design optimization framework and furnishes valuable insights into optimizing Gurney flap parameters. The comparison of metaheuristic algorithms serves to enhance the study's contribution to Gurney flap design optimization.
\end{abstract}


\begin{highlights}
\item The proposed optimization framework using the Cuckoo Search algorithm and Radial Basis Function neural network significantly improves the computational efficiency for designing Gurney flaps on airfoils.
\item The Cuckoo Search algorithm outperformed other state-of-the-art algorithms for optimizing the Gurney flap parameters.
\item The optimum Gurney flap configuration obtained after optimization improved the aerodynamic efficiency of the airfoil by 10.28\%.
\end{highlights}



\end{frontmatter}

\makenomenclature
\nomenclature{\(AoA\)}{angle of attack [\degree]}
\nomenclature{\(h\)}{Gurney flap height [mm]}
\nomenclature{\(N_p\)}{number of nests in population [-]}
\nomenclature{\(t\)}{current iteration [-]}
\nomenclature{\(T_{max}\)}{max iterations [-]}
\nomenclature{\(f(x)\)}{objective function [-]}
\nomenclature{\(x_i^t\)}{current solution of cuckoo i [-]}
\nomenclature{\(x*\)}{global optimum solution [-]}
\nomenclature{\(p_a\)}{probability of cuckoo egg being discovered and the corresponding nest being abandoned [-]}
\nomenclature{\(\alpha\)}{step size control scaling factor [-]}
\nomenclature{\(s\)}{random step size drawn from the Lévy distribution [-]}
\nomenclature{\(\beta_s\)}{step size control parameter [-]}
\nomenclature{\(\Gamma\)}{step size scaling factor [-]}
\nomenclature{\(R\)}{radius of flow domain [m]}
\nomenclature{\(I\)}{turbulent intensity [\%]}
\nomenclature{\(U_{\infty}\)}{freestream velocity [m/s]}
\nomenclature{\(Re\)}{Reynolds number [-]}
\nomenclature{\(A\)}{reference area for force non-dimensionalization [$m^2$]}
\nomenclature{\(l\)}{reference length for force non-dimensionalization [m]}
\nomenclature{\(h_c\)}{representative cell length [m]}
\nomenclature{\(A_p\)}{area of grid cells [m]}
\nomenclature{\(r\)}{refinement ratio [-]}
\nomenclature{\(p\)}{order of convergence [-]}
\nomenclature{\(e\)}{relative error [\%]}
\nomenclature{\(e^{extr}\)}{extrapolated relative error [\%]}
\nomenclature{\(GCI\)}{grid convergence index [\%]}
\nomenclature{\(N_c\)}{cell count [-]}
\printnomenclature

\section{Introduction}
\label{sec:sample1}


The economic viability of a commercial aircraft is heavily dependent on its high-lift performance. For a given wing size, the payload capacity and the flight range of an aircraft can be increased using an efficient high-lift system. Furthermore, improved lift performance enables steeper takeoff ascents, which can significantly reduce noise pollution around airports. By enhancing the efficiency of the wing during ascent, the aircraft can attain its optimal cruising altitude quickly, resulting in a more fuel-efficient flight. As a result of these benefits, aircraft designers and engineers have increasingly turned their attention to these devices over time. Trailing-edge flaps and leading-edge slotted flaps are examples of such devices currently being used in the aviation industry. The flow over these elements involves boundary layer interactions, main element wakes, flow through flap slot, and potential flow outside the boundary layer. These complexities make the design, manufacturing, and maintenance of these devices very challenging, providing an impetus for the implementation of mechanically simpler performance enhancement devices like Gurney flaps (GF).

A GF is a small rectangular strip-like element fixed near the airfoil's trailing edge facing the pressure side to enhance its performance. Its size varies from 1\% to 5\% of the chord (c) \citep{Storms1994}. It was invented by the American racecar driver, Dan Gurney who implemented it on the rear wing of his racecar. While conducting on-track testing he found that his racecar was able to carry more momentum in the corners while also attaining greater straight-line speeds \citep{Jang}, indicating an improvement in the ratio of lift ($C_l$) and drag ($C_d$) coefficients. Gurney flap operates by inducing a separation of the airflow as it passes over its rear end near the trailing edge causing a pressure decrement near the upper surface and an increment in pressure near the lower surface of the airfoil \citep{Myose1998}. This results in a decrement in the zero-lift angle of attack, an increment in the airfoil's maximum lift coefficient ($C_l$ max), and also an increment in the nose-down pitching moment ($C_m$), typically indicating an increment in the overall camber of the airfoil \citep{Myose1998}. It also leads to an increment in the value of $C_d$, particularly at lower angles of attack (AoA) \citep{Bloy1997}, but for thick airfoils, \citet{Neuhart1988AWT} has reported a reduction in the drag force. For obtaining a net improvement in $C_l$/$C_d$, the flap must be sized appropriately, taking into account the boundary layer thickness, as recommended by \citet{Giguere1997}.

Over time, numerous experiments and numerical simulations have been performed to understand flow physics and optimize the performance of Gurney flaps. \citet{Liebeck1978} conducted an experimental study on Newman airfoil using a GF having a flap height (h) of 1.25\%c. The results showed an increment in the lift force (L) and a small decrement in the drag force (D) acting on the airfoil. Also, it was observed that larger flap heights resulted in a greater increase in lift, but this was accompanied by a significant increase in drag beyond flap heights of 2\%c. To explain these effects, \citet{Liebeck1978} hypothesized that the reversed flow region behind a GF comprises two contra-rotating vortices, altering the circulation and the Kutta condition in that region. \citet{Katz1989}, as well as \citet{Katz1989_2}, later conducted wind tunnel experiments by incorporating a GF on the trailing element of multi-element racecar wings. Their studies also reported a significant increment in lift, with the use of a 5\%c GF resulting in up to a 50\% lift increase. However, the corresponding $C_l$/$C_d$ of the wing was found to decrease over the design angle of attack range $(2\degree \leq AoA \leq 12\degree)$.

\citet{Jang} conducted numerical investigations on NACA 4412 airfoil with GF attached to its trailing edge. The simulated GF heights (h) ranged from 0.5\%c to 3\%c. The numerical solution indicated a significant increase in $C_l$ and $C_m$ with the use of GF. An increment in $C_l$ with a slight increase in $C_d$ was observed with flap heights less than 1.25\%c. Compared to the baseline case (without the GF), the computed pressure distribution indicated an increment in the load along the airfoil section, predominantly near the trailing edge and the suction peak. Also, with the use of the GF, \citet{Jang} reported a shift in the separation point farther aft at moderate angles of attack compared to the clean airfoil. These numerical results were consistent with the wind tunnel tests performed by \citet{Wadcock1987}.

A comprehensive experimental study was performed by \citet{Jeffrey2000} on single-element wings equipped with a Gurney flap. Forces, surface pressures, and velocities were measured using Laser Doppler Anemometry (LDA). The temporal-averaged streamlines and velocity vectors showed the presence of contra-rotating vortices behind the GF, which was in line with the earlier hypothesis by \citet{Liebeck1978} and numerical data of \citet{Jang}.  Additionally, smoke visualizations and LDA measurements revealed the presence of a Karman vortex street in the wake region of the GF. The periodic vortex shedding decreases the trailing edge pressure, while the bluff shape of GF retards the flow at the windward side of GF, increasing the pressure on the lower surface, thus increasing the overall lift force.
\citet{Bloy1995} carried out wind tunnel tests for finding surface pressure distributions and wake profiles of NACA $63_2-215$ airfoil having a GF and a $45\degree$ trailing edge flap. The data gathered from the experiments indicated that the $45\degree$ trailing edge flap performed better in comparison to the GF for the same flap height. Significant lift increments were observed for both flaps, with the trailing edge flap providing a lesser drag penalty compared with the GF. The addition of a 2\%c $45\degree$ trailing edge flap resulted in a significant increase in $C_l$ max, raising it from 1.06 to approximately 1.40. The maximum $C_l/C_d$ remained comparable to the baseline airfoil without GF but was achieved at a higher value of $C_l$. Furthermore, the flapped airfoil produced a lower value of $C_l/C_d$ at low to moderate values of $C_l$.

Hence, it can be concluded that the performance of the GF is highly sensitive to the values of its design parameters (flap height, width, mounting angle, and position on the chord line), which can be challenging to determine through traditional methods. To address this challenge, we propose an optimization framework that uses a metaheuristic algorithm called the Cuckoo Search (CS) algorithm \citet{CSorg} for finding the optimal parameters for the GF. Studies \citep{AES1,AES2,AES3} in the past have shown promising results and the CS algorithm has been shown to outperform widely used algorithms like PSO and GA on several occasions \citep{CScomp1,CScomp2,CScomp3}. Previously, the Cuckoo Search algorithm has been utilized to address a broad spectrum of optimization problems. \citet{ecoli} used the CS algorithm in an E. Coli fed-batch fermentation procedure to identify the model parameters for the process. Their results indicated that the CS algorithm was highly effective in identifying optimal parameters for the process when compared to Genetic Algorithm (GA) and Ant Colony Optimization (ACO). \citet{yildiz} implemented the CS algorithm for optimizing machining parameters in milling operations. The results of the study indicated a substantial reduction in manufacturing time when compared to other optimization techniques like ACO, PSO, and handbook recommendations. \citet{kaveh} highlights the advantages of using the CS algorithm for designing two-dimensional steel frames. The findings indicated that Lévy flights were more efficient and effective in the searching process for large design spaces compared to the Gaussian or uniform distributions commonly used by other metaheuristic algorithms. \citet{valian} used a modified CS algorithm where the parameters are tuned to increase the algorithm’s efficiency and used the improved CS algorithm by calculating the weights and biases of a neural network. Their study highlights the importance of metaheuristic algorithms in neural network parameter optimization problems where a differentiable neuron transfer function is not available. 

Studies in the past have either analyzed only one parameter of the GF \citep{GF_single} or used time-consuming heuristic methods to optimize the multiple GF parameters \citep{Li2003}. In the present study, a novel optimization framework is employed to optimize the mounting angle ($\theta$) and height (h) of a Gurney flap. The present study can be broken down into three objectives:
\begin{enumerate}
    \item The first objective is to develop a framework comprising the usage of CFD data to train a surrogate model.
    \item The second objective is to execute the CS algorithm to obtain the Gurney flap configuration with the optimal design parameters and analyze its effect on the flow field.
    \item The third objective is to quantitatively compare the CS algorithm with respect to five other prominent metaheuristic algorithms. 
\end{enumerate} 


\begin{figure}
    \centering
    \includegraphics[width=150mm,scale=0.8]{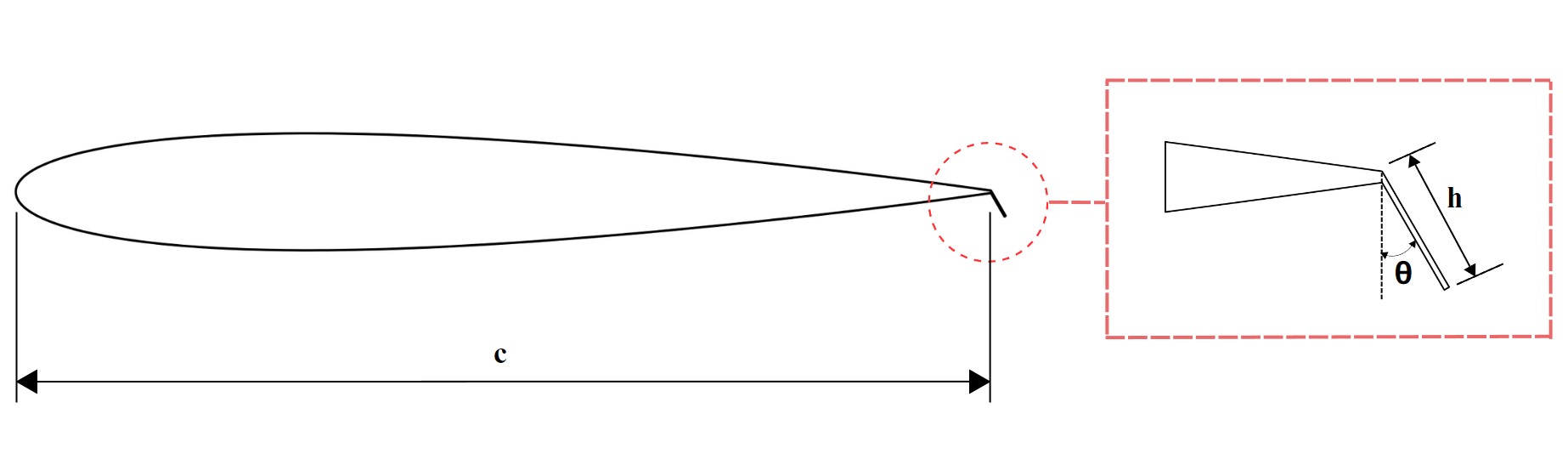}
    \caption{Schematic of NACA 0012 airfoil with GF fixed at the trailing edge}
    \label{schematic}
\end{figure}

\section{Computational Modeling}
\label{Sec2}
\subsection{Gurney Flap Layout and Parameterization}

NACA 0012 airfoil of chord (c) 1m was chosen for the optimization study. This choice was based on the comprehensive experimental research conducted by \citet{Li2002}, which extensively examined the baseline as well as the airfoil with a GF. The configuration of the GF is determined by four key parameters: flap height, flap width, mounting angle in relation to the chord, and its position along the chord. Figure \ref{schematic} depicts the schematic of a NACA 0012 airfoil having a GF mounted to its trailing edge facing the pressure side. The flap height is represented by h, and the angle made by the GF with the normal to the chord is denoted by $\theta$.
The flap angle is assigned a positive value when measured clockwise and a negative value when measured anticlockwise from the line normal to the chord. 
Based on past experimental and numerical studies \citep{Storms1994,Bloy1997,Neuhart1988AWT,Giguere1997,Liebeck1978,Li2002}, adjusting h and $\theta$ of the GF can significantly improve the airfoil's efficiency ($C_l$/$C_d$), while changing the GF width has no significant impact on the airfoil's performance. Additionally, investigations conducted by \citet{Li2003} on NACA 0012 to examine the impact of the mounting position of GF on the airfoil's performance showed that mounting the GF at the trailing edge resulted in a more efficient airfoil than shifting it forward. Therefore, the present optimization study focuses on the parameters, h and $\theta$, with GF placed at the trailing edge and the flap width kept constant at 0.2\%c. The limits for both parameters are determined based on previous investigations, with h ranging from 0.5\%c to 4\%c and $\theta$ ranging from $-60 \degree$ to $30 \degree$.

\subsection{Numerical Model}
The surrogate model for the optimization study was trained on the CFD data generated using two-dimensional RANS simulations carried out using ANSYS Fluent version 22.1. The selection of a turbulence closure model is a crucial part affecting the accuracy of CFD solutions. The flow over an airfoil involves adverse pressure gradients, flow separation, and recirculation zones, hence the $k-\omega$ Shear Stress Transport (SST) turbulence model of \citet{Menter1993} was used to carry out these simulations due to its ability to model such type flows accurately, which was verified by past numerical studies on airfoils \citep{Kral1998,Yu2011,Zhang2009}. 
It utilizes a blending function to switch between Wilcox $k-\omega$ and $k-\epsilon$ model. Close to a wall, the blending function preserves the characteristics of the $k-\omega$ model while gradually transitioning towards the conventional $k-\epsilon$ model as the distance from the surface increases. This is done due to the robustness of the $k-\omega$ model in capturing the effect of the viscous sublayer and using the $k-\epsilon$ model in the freestream region to overcome the sensitivity of the $k-\omega$ turbulence model to entrance turbulence parameters.

\subsection{Computational Grid}
\begin{figure}[h]
\centering
  \includegraphics[width=120mm]{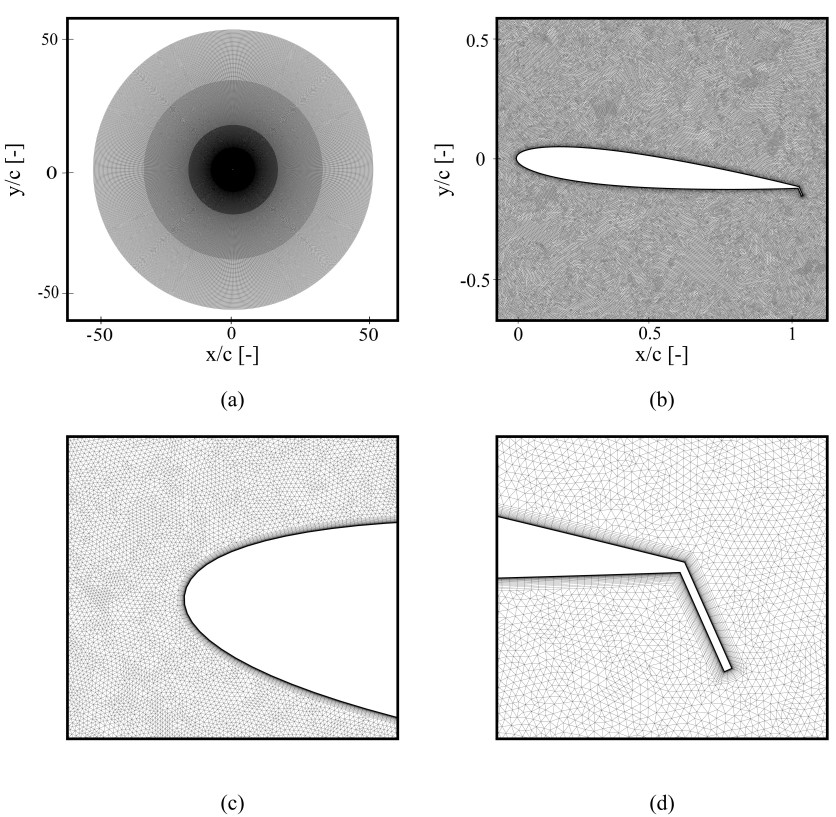}
\caption{Computational grid of NACA 0012 airfoil (AoA=6$\degree$) equipped with a GF (h=30\%c and $\theta$=-30\degree). (a) Overall mesh, enlarged views near the (b) airfoil, (c) leading edge, and (d) trailing edge}
\label{domain}
\end{figure}

An O-type flow domain was used for the CFD simulations. The chord of the NACA 0012 airfoil was taken as 1m. The origin of the flow domain was fixed at the leading edge of the airfoil and the outer periphery of the domain was kept at 50 times the chord, $R=50c$ from the origin. A large domain size was chosen to prevent flow reflection from the outer periphery. Furthermore, the outer periphery was split into two parts, where the left part served as the inlet and the right part as the outlet. To reduce the discretization error, second-order grid cells were used. The flow domain was divided into two parts- the region close to the airfoil having an unstructured tri mesh and the other part having a structured O-grid quad mesh. This was done to avoid bad quality and high aspect ratio cells near the airfoil and the GF. Another major reason for taking this approach is based on the parametric framework that was developed for carrying out the simulations. In simulating each new design point (with a new value of h and $\theta$ for the GF) the geometry would get updated first, followed by the remeshing and solution process. By taking the current approach of splitting the domain, the mesh generator would keep the mesh in the outer domain unchanged while updating the mesh only inside the inner domain. This speeds up the meshing process and ultimately reduces the overall simulation time. The $y^+$  of the grid was kept below 1 to capture the boundary layer effects and about 32 inflation layers were wrapped around the airfoil and the GF. The first layer height was taken as $5\times10^{-3} mm$, and the height of subsequent layers was based on a constant growth rate of 1.14. Figure \ref{domain} shows the cell distribution inside the flow domain of the airfoil equipped with a GF ($h=30mm$ and $\theta=-30\degree$). A higher grid density near the airfoil and the GF can be observed from Figure \ref{domain}. This was done to resolve the flow physics accurately in regions of the flow domain with high-velocity gradients and turbulence and at the same time keep the overall computational cost low by avoiding excessive cells in far-field regions.

\subsection{Setup and Boundary Conditions}
Steady-state RANS simulations were carried out utilizing k-$\omega$ SST \citep{Menter1993} model for turbulence closure. The inlet was assigned with the default values of turbulent intensity (I=5\%) and turbulent viscosity ratio ($\mu_t/\mu= 10$). A freestream velocity ($U_{\infty}$) of 31 m/s (Mach number= 0.091) was assigned at the inlet and the Reynolds number (Re) was kept at  $2.1\times10^6$. The no-slip condition was assigned to the airfoil and the outlet was assigned the pressure outlet boundary condition. The operating pressure was taken as 101325 Pa, and both the inlet and outlet were assigned a gauge pressure of 0 Pa. Since the Mach number for the simulation is much less than 0.3, an incompressible pressure-based solver was used for the simulation. The density ($\rho=1.225 kg/m^3$) and the dynamic viscosity ($\mu=1.7894\times10^{-5} kg/m.s$) were kept constant. The force coefficients ($C_l$ and $C_d$) were non-dimensionalized using a reference area, $A=1.6 m^2$, and a reference length, $l=1m$.
The pressure-velocity coupling utilized the COUPLED algorithm, while gradient discretization employed the least square cell-based scheme. For the spatial discretization of pressure, momentum, turbulent kinetic energy, and specific dissipation rate, the second-order linear upwind scheme was used. To ensure the convergence of $C_l$ and $C_d$ monitors and the residuals to fall below the specified limit of $1\times10^{-5}$, 1000 iterations were used for all the simulation cases.

\subsection{Grid Convergence Study and Numerical Model Validation}
To examine the sensitivity of the CFD solution on the grid size and estimate the discretization error, a grid convergence study was conducted using the Richardson Extrapolation method \citep{procedureNew}. For the study, three different grids were constructed, namely Coarse, Medium, and Fine and their cell counts are $2\times10^5$, $4\times10^5$, and $8\times10^5$ respectively. 
For clarity and ease of tracking, instead of cell count $N_c$, the representative cell length $h_c$ was used for the study as  $h_c\to 0$ while $N_c \to \infty$ for an infinitely fine grid. The representative cell length  $h_c$ is given by

\begin{equation}
    h_c = \frac{1}{N_c}\sum_{Cells}A_p^{1/2}
\end{equation}

\begin{figure}[h]
 
\begin{tabular}{cc}
  \includegraphics[width=65mm]{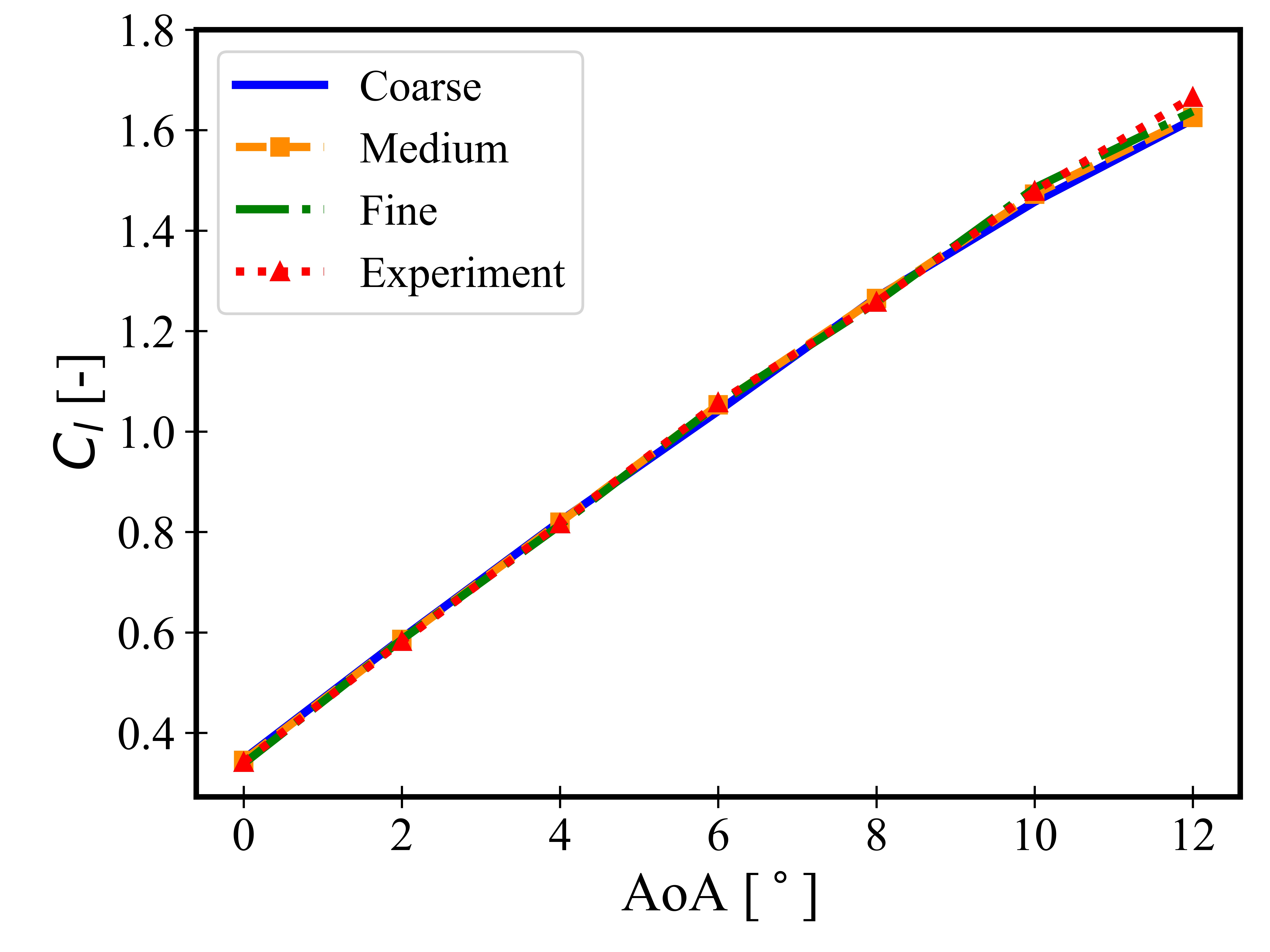} &   \includegraphics[width=65mm]{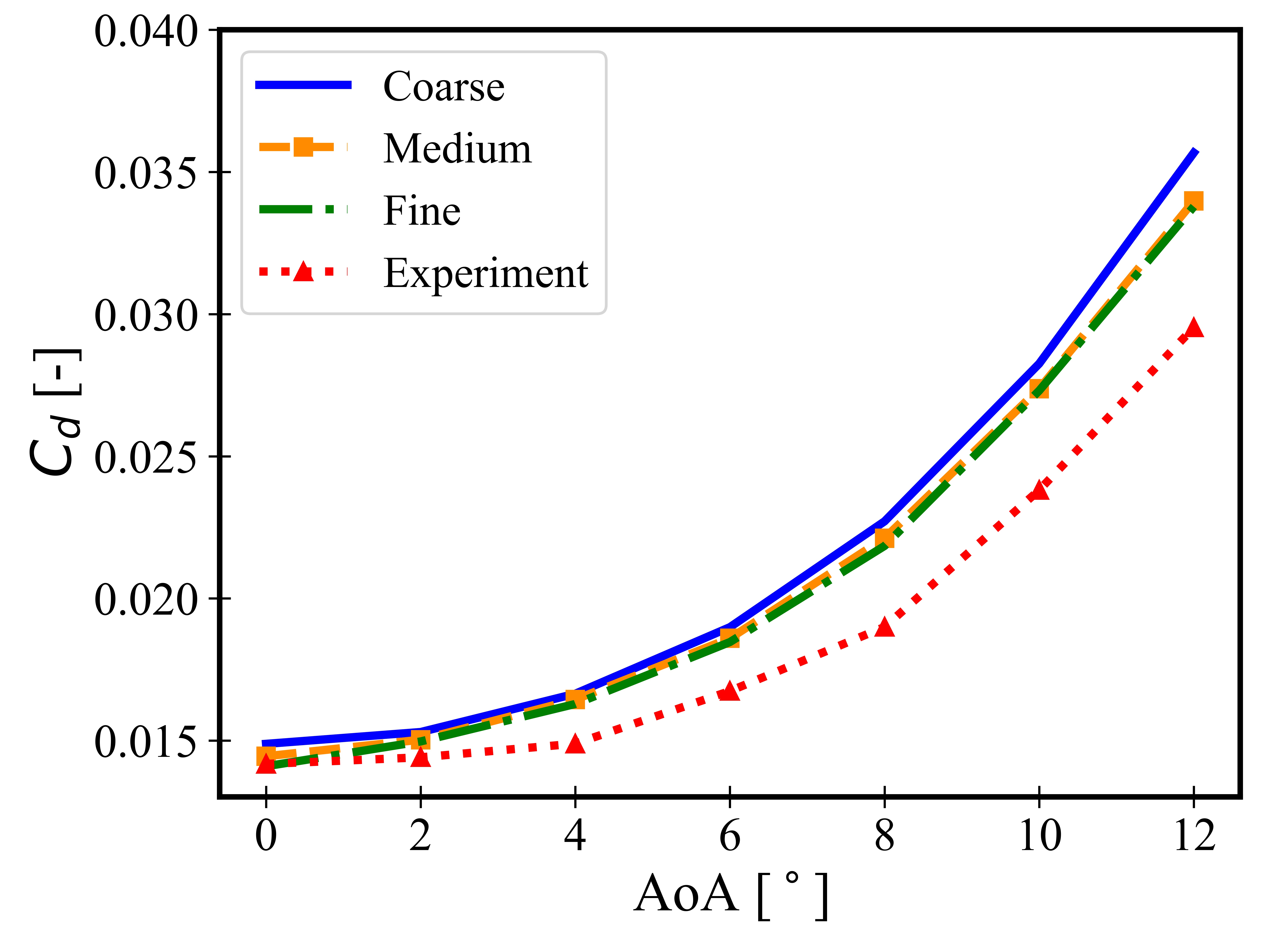} \\
(a) & (b) \\[6pt]
\end{tabular}
\caption{Comparison of CFD results for three different grid resolutions (Coarse, Medium, and Fine) with experimental \citep{Li2002} data for the airfoil with GF (h=2\%, $\theta=0\degree$). (a) $C_l$ and (b) $C_d$}
\label{flapFig}
\end{figure}

\begin{figure}[h]
 
\begin{tabular}{cc}
  \includegraphics[width=65mm]{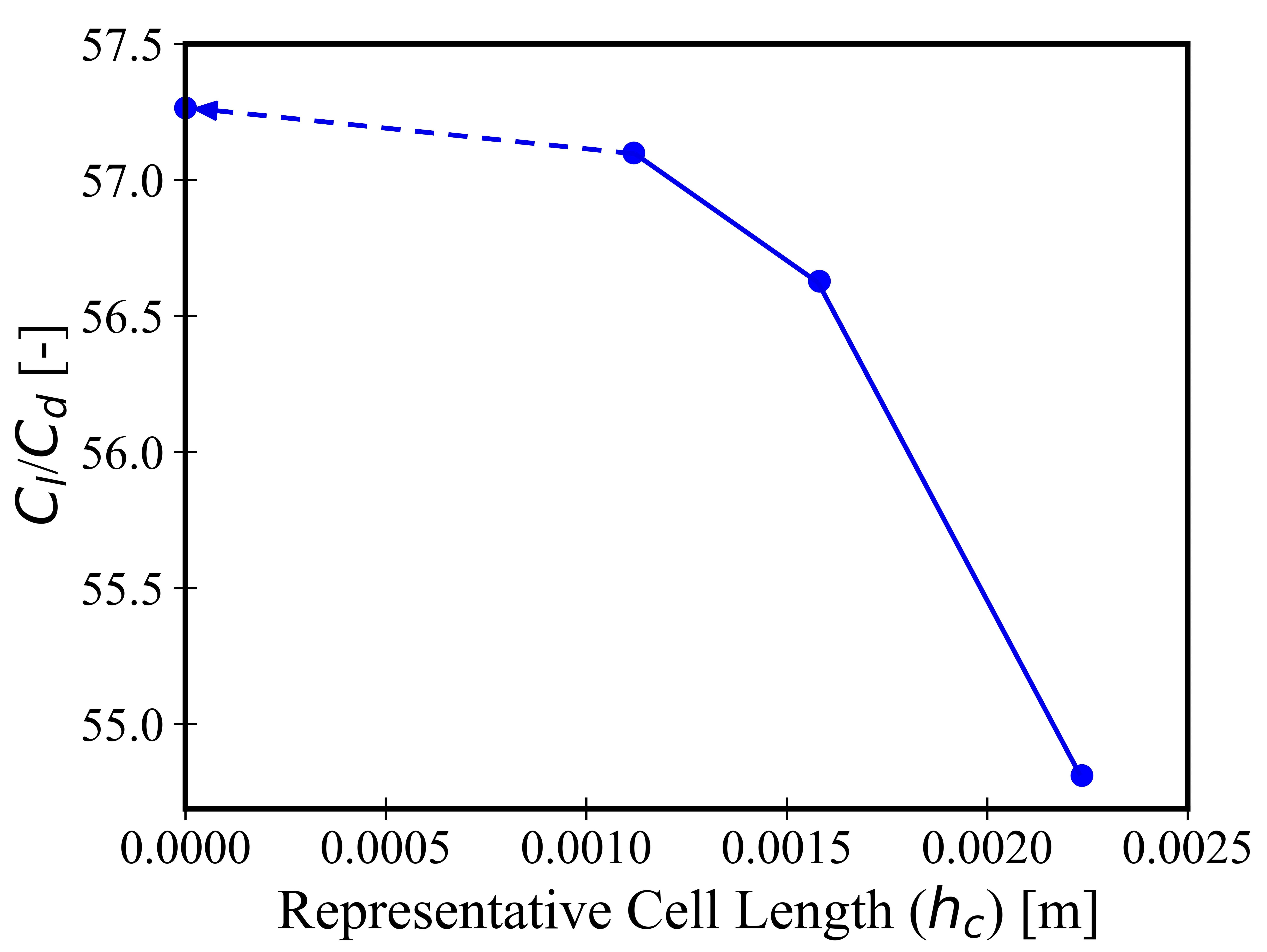} &
  \includegraphics[width=65mm]{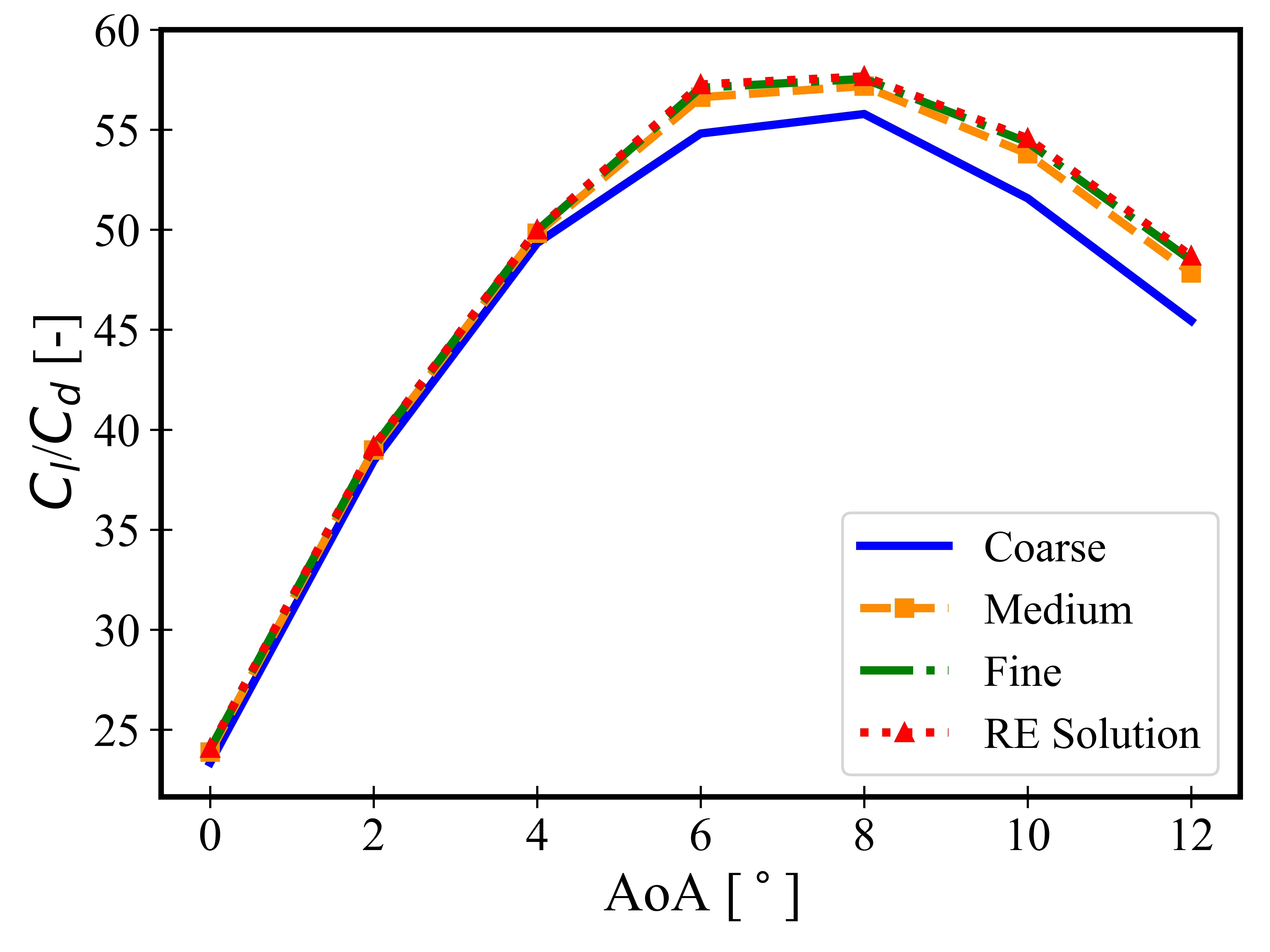} \\
  (a) & (b) \\[6pt]
\end{tabular}
\caption{(a) Variation of the computed value of $C_l/C_d$ for the airfoil with GF (h=2\%, $\theta=0\degree$) at AoA=$6\degree$ with decreasing grid sizes and (b) Comparison of $C_l/C_d$ for the airfoil with GF (h=2\%, $\theta=0\degree$) computed using three different grid resolutions (Coarse, Medium, and Fine) with Richardson Extrapolation (RE) solution}
\label{flapRE}
\end{figure}

where $A_p$ is the cell area and $N_c$ is the cell count in the grid.
After this, the refinement ratio $r$ was calculated which is the ratio of $h_c$ for medium and fine grids. \citet{procedureNew} recommends that the representative cell lengths (h) should be at least 30\% different between each grid size. In the next step, the order of convergence $p$ of the grid was determined based on the values of solution variable $\phi$ for the coarse, medium, and fine grids. For the present study, the ratio of $C_l/C_d$ was selected as the solution variable. The parameter $p$ gives a quantitative indication of the change in discretization error with a change in grid size.
The order of convergence $p$ is given by

\begin{equation}
    p = \frac{\ln \left( \frac{\phi_{coarse} - \phi_{medium}}{\phi_{medium} - \phi_{fine}}\right)}{\ln(r)}
\end{equation}

After finding the parameter $p$, the solution variable for an infinitely fine grid $(N_c \to \infty$  or  $h_c \to 0)$ $\phi_{exact}$ was estimated. It is also called Richardson Extrapolation (RE) solution and is given by

\begin{equation}
    \phi_{exact} = \phi_{fine} - \frac{\phi_{medium} - \phi_{fine}}{r^p - 1}
\end{equation}

From all the values calculated above, three different error metrics are calculated. 
These are as follows:
\begin{enumerate}
    \item Relative Error: It indicates the closeness of the solution
    of the medium and fine grids. It is not the most reliable metric since it ignores the estimated solution of the infinitely fine grid. It is given by
    \begin{equation}
        e = \frac{\phi_{medium} - \phi_{fine}}{\phi_{fine}}
    \end{equation}
    \item Extrapolated Relative Error: It indicates the closeness of the solution of fine grid to the estimated value for the infinitely fine grid. It is given by
    \begin{equation}
        e^{extr} = \frac{\phi_{fine} - \phi_{exact}}{\phi_{exact}}
    \end{equation}
    \item Grid Convergence Index: It was proposed by Roache \citep{perspective} for improving the relative error metric. The relative error is scaled by a factor of $1/(r^p - 1)$. This scaling accounts for the order of convergence of the solution, which is absent in the definition of relative error. To complete its definition, Roache \citep{perspective} applied a factor of safety of 1.25. Hence, it is given by
    \begin{equation}
        GCI = \frac{1.25e}{r^p - 1}
    \end{equation}
\end{enumerate}

The findings of the grid convergence and validation studies are summarized below:
\\For the validation of the numerical model, results of wind tunnel experiments carried out by \citet{Li2002} were used. 
Figure \ref{flapFig} represents the comparison of the CFD solutions with the wind tunnel data for $C_l$ and $C_d$. At higher values of AoA, the discrepancy between the CFD results and experimental data grows more pronounced, particularly when using coarser grid resolutions. By using the fine grid, the maximum deviation from the experimental results for the airfoil equipped with GF (h=2\%, $\theta=0\degree$) was found to be 1.69\% and 15.00\% for $C_l$ and $C_d$, respectively.
The deviation in the results could be due to various factors: geometric simplification and 2D modeling in CFD simulations, differences in the surface roughness characteristics of the airfoil, uncertainties in measurements inside the wind tunnel, differences in freestream turbulence parameters, steady state assumption, numerical and modeling errors, etc. Despite this, the deviation in the results is within the acceptable range. Hence the fine grid will be used for the further part of the optimization study.

The solution at $h_c=0$ in Figure \ref{flapRE}(a) represents the RE solution or the solution at the infinitely fine grid. It also depicts the variation in the value of $C_l/C_d$ with decreasing grid sizes for the airfoil at AoA $6\degree$. It is evident that the solution obtained from the refined grid closely aligns with the RE solution. In addition, Figure \ref{flapRE}(b) represents the comparison of the computed solution for the coarse, medium, and fine grids with the RE solution at various AoA. This plot also shows the accuracy of the fine grid in capturing the flow physics better in comparison to the medium and coarse grids. Finally, computed values of all the error metrics discussed earlier and the order of convergence of the grids for the airfoil with GF are presented in Table \ref{table_GF}. It can be concluded that all the error metrics are less than 1.3\%  for the validation case. Also, the order of convergence of the grid is slightly less than two. This could be due to the presence of skewness errors in the grids and gradient limiters in the face interpolation schemes.

\begin{table}
\resizebox{\textwidth}{!}{%
    \begin{tabular}{ c c c c c c c c c}
     \hline
     \multirow{2}{3em}{AoA[$^{\circ}$]} & \multicolumn{4}{ c }{$C_l/C_d$ $[-]$} & \multirow{2}{2em}{p$[-]$} & \multirow{2}{1em}{e\%} & \multirow{2}{3em}{$e^{extr}[\%]$} & \multirow{2}{3em}{GCI[\%]}\\
      & Coarse & Medium & Fine& RE Solution & & & &\\
     \hline
     0 & 23.351 & 23.877 & 24.024 & 24.081 & 1.843 & 0.610 & 0.235 & 0.295\\
     2 & 38.446 & 38.987 & 39.124 & 39.170 & 1.984 & 0.350 & 0.118 & 0.148\\
     4 & 49.333 & 49.827 & 49.958 & 50.006 & 1.910 & 0.263 & 0.095 & 0.119\\
     6 & 54.811 & 56.628 & 57.100 & 57.265 & 1.945 & 0.826 & 0.289 & 0.362\\
     8 & 55.787 & 57.185 & 57.543 & 57.667 & 1.965 & 0.623 & 0.214 & 0.268\\
     10 & 51.571 & 53.798 & 54.373 & 54.573 & 1.953 & 1.058 & 0.367 & 0.460\\
     12 & 45.444 & 47.840 & 48.469 & 48.693 & 1.930 & 1.298 & 0.460 & 0.577\\
     \hline
    \end{tabular}}
\caption{Results of Grid Dependency Study for airfoil equipped with GF (h=2\%, $\theta=0\degree$)}

\label{table_GF}
\end{table}


\section{Optimization Framework}
\label{sec3}
\subsection{Sample Acquisition}
Before creating a successful surrogate model, an important step is to generate the required training data.
The Data points must be spread evenly to represent the entire design space. Latin Hypercube Sampling (LHS) by \citet{LHS} is a popular method for sampling data points in the design space.

\subsection{Surrogate Modeling}
Design optimization problems that make use of a large number of computational simulations require expensive computational resources \citep{wang2007}. Surrogate models may solve the problem of simulations' high computing costs. These models build a response surface from a small number of simulations to efficiently explore the design space and predict output variables. This enables researchers to simulate complicated systems more accurately with fewer computational resources. These models are trained using a subset of simulation data and validated using the rest. In previous investigations\citep{JCSRBF,MPA,AESRBF,nedelkova}, the RBF surrogate model has accurately described significant nonlinear interactions between variables. Its simplicity, ease of implementation, versatility, and efficiency made it the appropriate surrogate model for this investigation.\\

\subsection{Cuckoo Search Algorithm}
Cuckoo Search Algorithm was proposed by \citet{CSorg}. The algorithm takes its inspiration from the behaviour of cuckoos and their brood parasitism. Cuckoo birds exhibit a highly aggressive reproductive strategy, with certain species laying eggs in communal nests and removing the host bird's eggs to increase the chances of their eggs being hatched. As a result, host birds often engage in conflict with cuckoos, with some host birds discovering the cuckoo eggs and responding by either abandoning the nest or building a new nest somewhere else. The probability of this behavior occurring is controlled by a parameter, denoted by $p_a$.
Studies have also shown that cuckoo chicks can mimic host chicks' calls, potentially gaining more opportunities for feeding from host parents. These adaptive behaviors suggest that cuckoos have evolved to exploit the reproductive strategies of their host birds, allowing them to increase their own fitness and ensure their survival.

\begin{algorithm}[h]
\setstretch{1}
\SetAlgoLined
\KwIn{Number of nests $N$, maximum generation $T_{\max}$, fitness function $f(\cdot)$}
\KwOut{Global optimum solution $\mathbf{x}^\star$}
Objective Function $f(x)$ \par
Initialize nests with random solutions $\mathbf{x}_i$ $ (i=1,2,.., N)$ and set $t\gets 0$\par
\While{$t < T_{max}$}{
    \hskip\algorithmicindent Get random cuckoo $i$ and use Lévy  flights for generating new solution;\par
    \hskip\algorithmicindent Calculate fitness of each nest $f_i=f(\mathbf{x}_i)$\par
    \hskip\algorithmicindent Sort the nests based on their fitness values.\par
    \hskip\algorithmicindent Randomly select a nest (say  $j$);\par
    \If{$f(\mathbf{x_i})>f(\mathbf{x_j})$}{
        Replace $j$ with the new solution
    }
    \hskip\algorithmicindent Abandon a fraction ($P_a$) of worse nests \par
    \hskip\algorithmicindent Build new nests via Lévy  flights \par
    \hskip\algorithmicindent Nests with best solutions are kept \par
    \hskip\algorithmicindent Rank the solutions and find the current best \par
    
}
\hskip\algorithmicindent $t\gets t+1$ \par
\Return{$\mathbf{x}^\star$}\;
\caption{Cuckoo search algorithm}
\label{algo}
\end{algorithm}

\begin{figure}[h]
    \centering
    \includegraphics[width=150mm,scale=0.8]{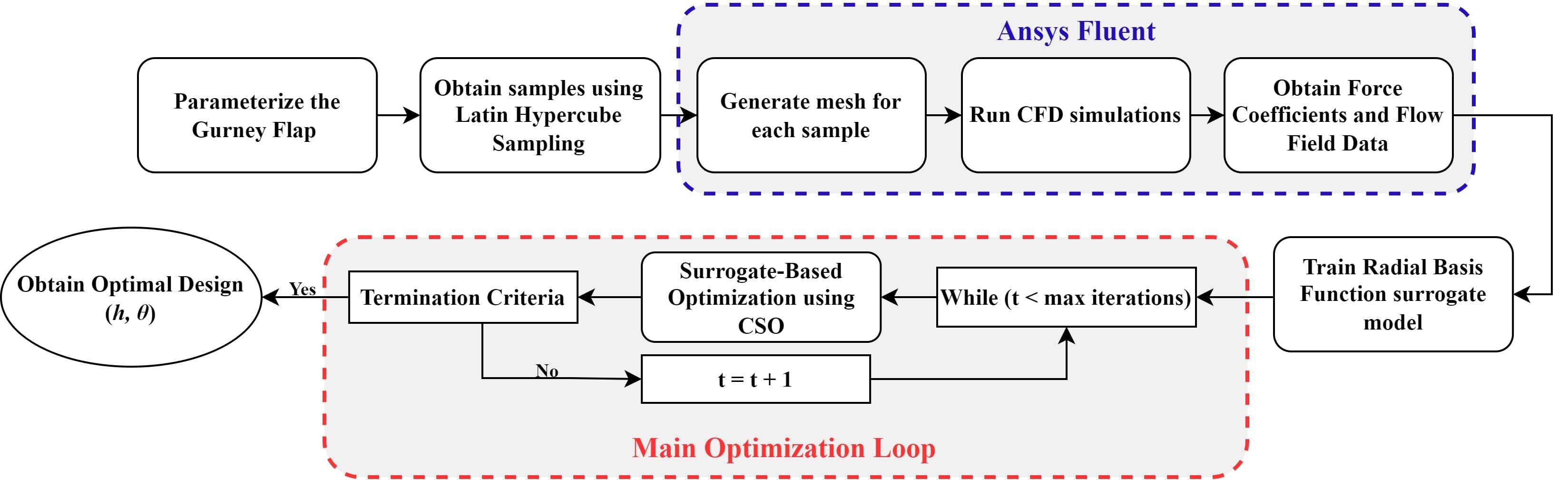}
    \caption{Flowchart of Optimization Framework}
    \label{flowchart}
\end{figure}



\begin{figure}[h]
\centering
  \includegraphics[width=65mm]{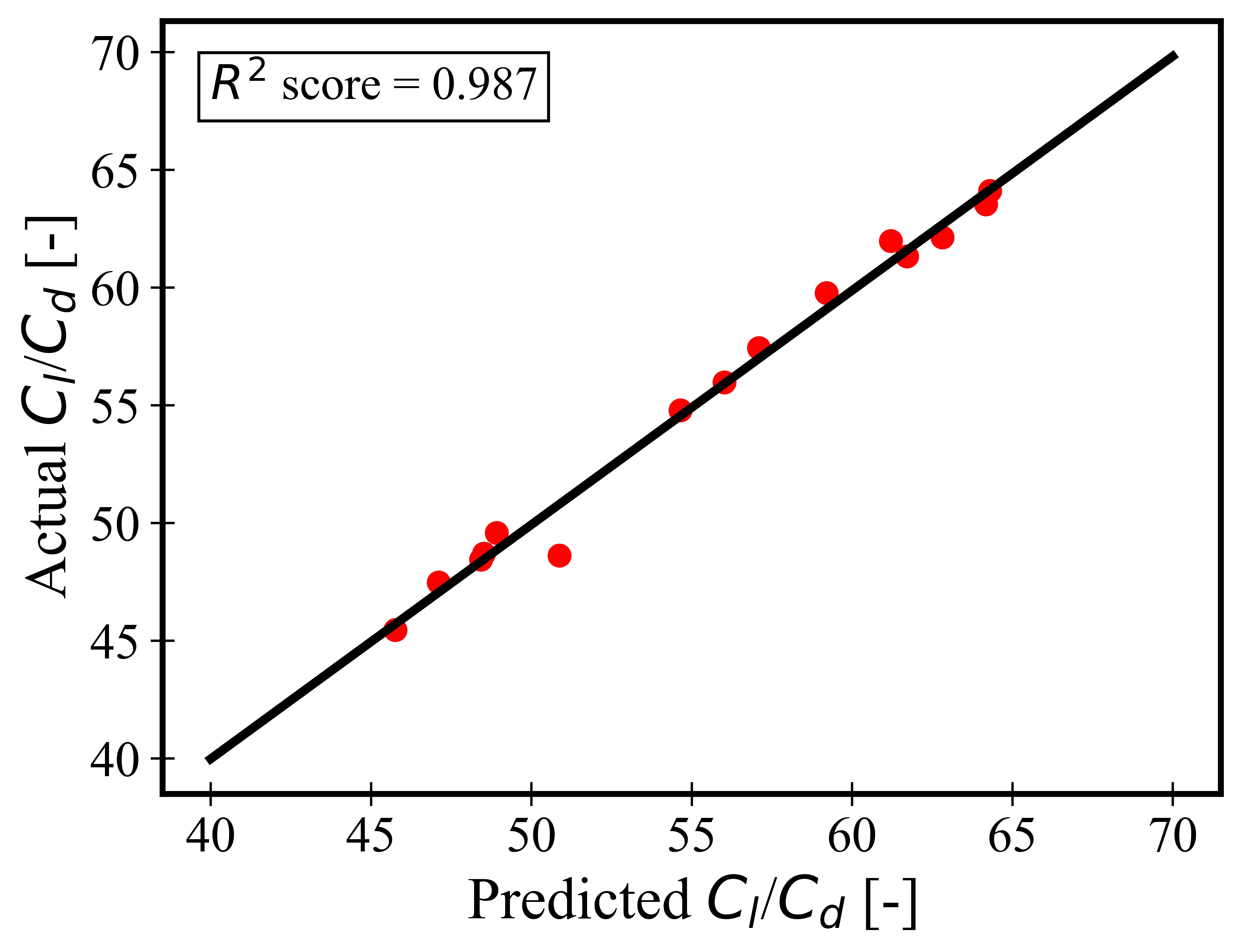}
\caption{Validation of the RBF surrogate model}
\label{r2_score}
\end{figure}

For maximization problems, such as the one in the present study, the quality of an egg is proportional to $C_l/C_d$ at each design point. Algorithm \ref{algo} presents the pseudo-code of the CS algorithm.
After successful training of the surrogate model and the subsequent generation of the response surface, the CS algorithm is used on the response surface to obtain the most optimal configuration of the gurney flap. The entire framework is summarized in a flowchart shown in Figure \ref{flowchart}.

\section{Results and Discussion}
\subsection{Radial Basis Function Surrogate Model}
\begin{figure}[h]
 
\begin{tabular}{cc}
  \includegraphics[width=65mm]{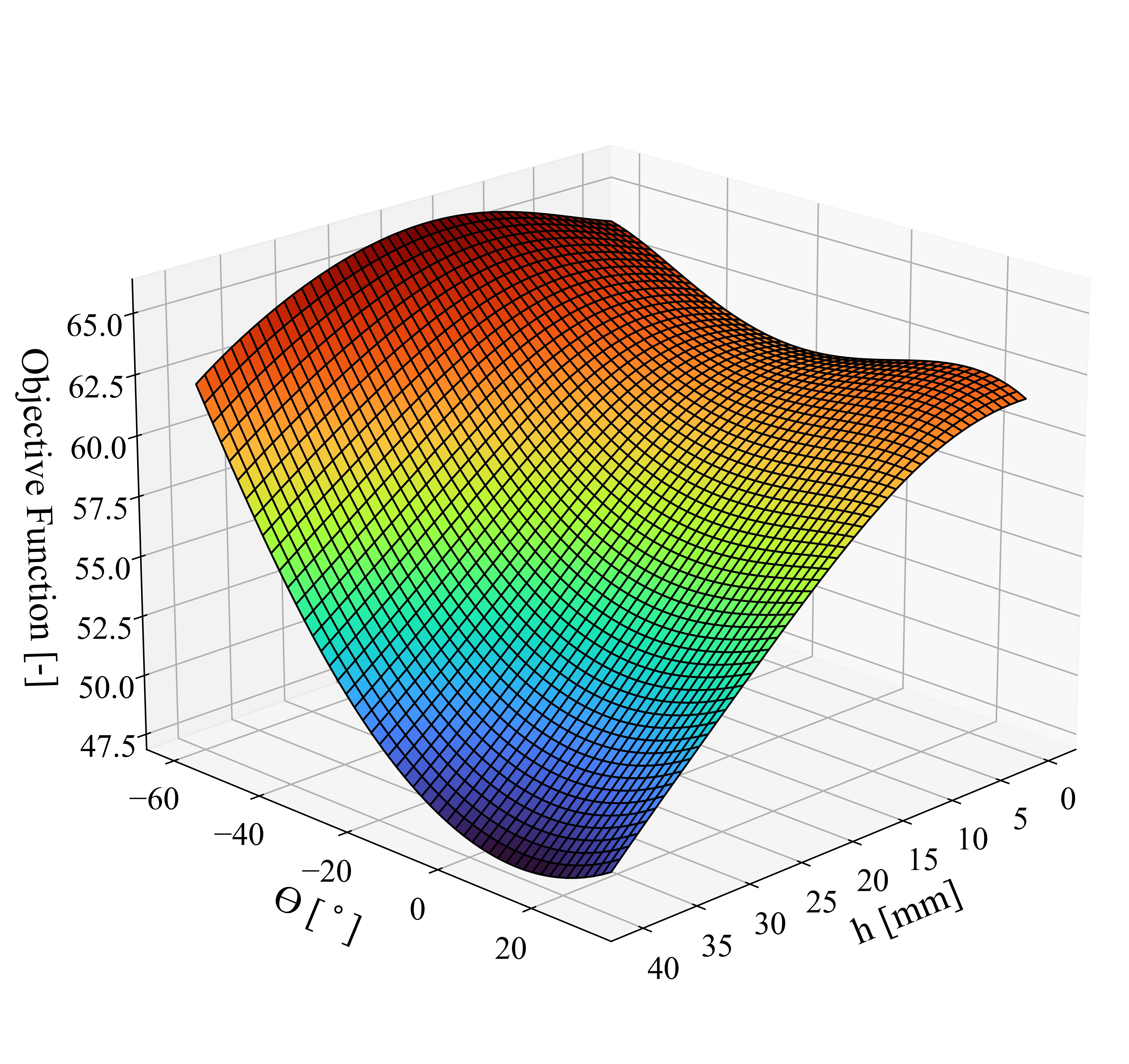} &   \includegraphics[width=65mm]{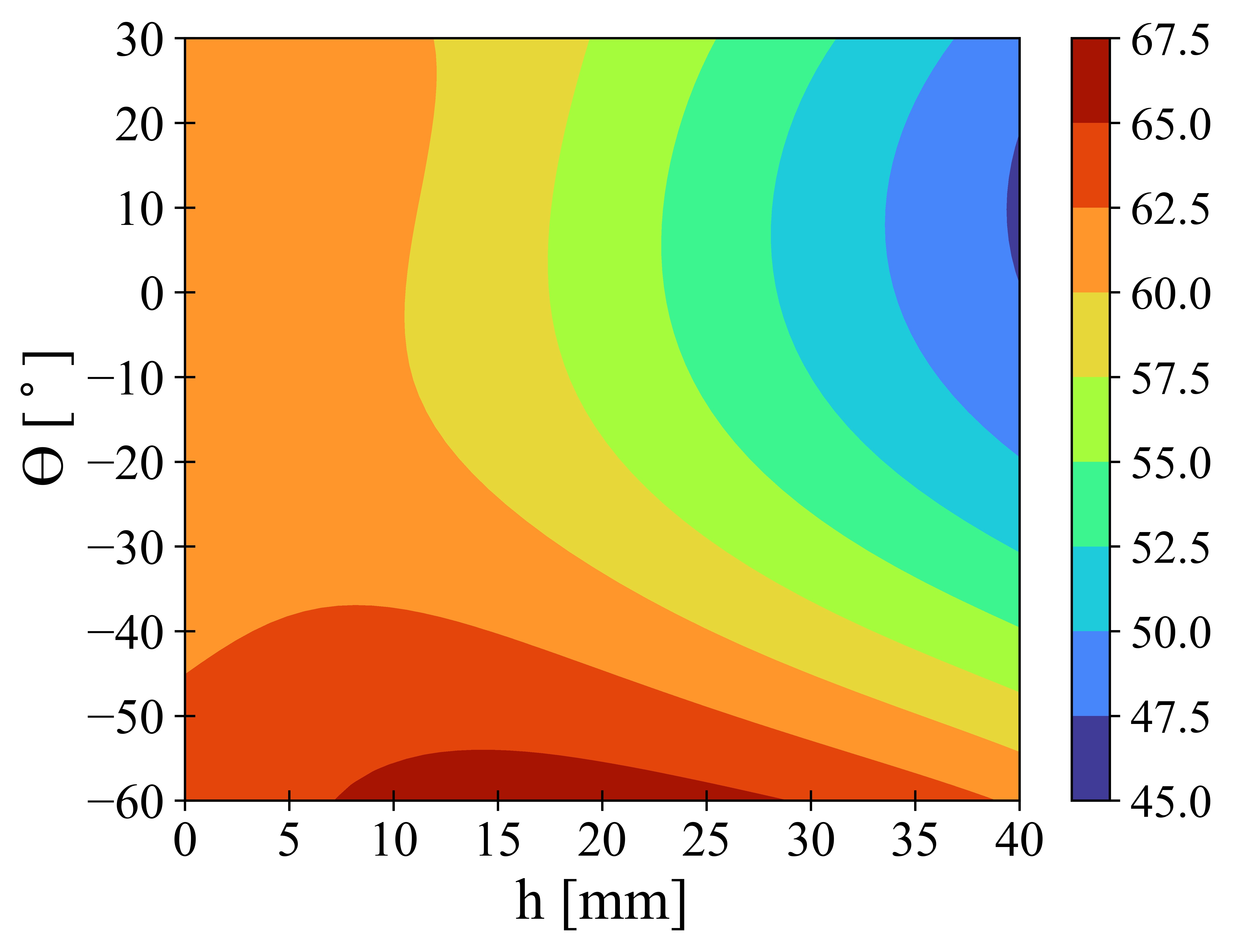} \\
(a) & (b) \\[6pt]
\end{tabular}
\caption{Response surface generated using RBF surrogate model}
\label{responseSurface}
\end{figure}
80 sample points with different design parameters were generated using LHS, and these sample points were used to carry out CFD simulations. Out of the 80 sample points, 65 were used for training the surrogate model, and the rest were used to validate the model. In light of minimal deviation from experimental data, an AOA of $6\degree$ was selected as the focal point of our study. While a similar framework could be constructed for other angles of attack, such an attempt was deemed redundant for the present study, and thus our analysis was concentrated on the aforementioned $6\degree$ AoA.
The accuracy of the RBF surrogate model has been assessed as shown in Figure \ref{r2_score}. The $R^2$ score of the data is observed to be 0.984.
The response surface generated by the surrogate model is shown in Figure \ref{responseSurface}, illustrating the strong non-linear dependence of the objective function on the design parameters. The response surface reveals that increasing h beyond 2.5\%c and increasing the flap mounting angle beyond -40$\degree$ results in a decrease in the airfoil efficiency ($C_l/C_d$). Conversely, the maximum airfoil efficiency is achieved when h is in the range of 1.5\%c to 25\%c, and the flap mounting angle is in the range of -60$\degree$ to -50$\degree$. These findings provide valuable insights for optimizing the design of Gurney flaps, as they provide an indication of the ranges of flap height and mounting angle that can significantly improve the airfoil efficiency.

\subsection{Cuckoo Search Probability Parameter}

\begin{figure}[h]
 
\begin{tabular}{cc}
  \includegraphics[width=65mm]{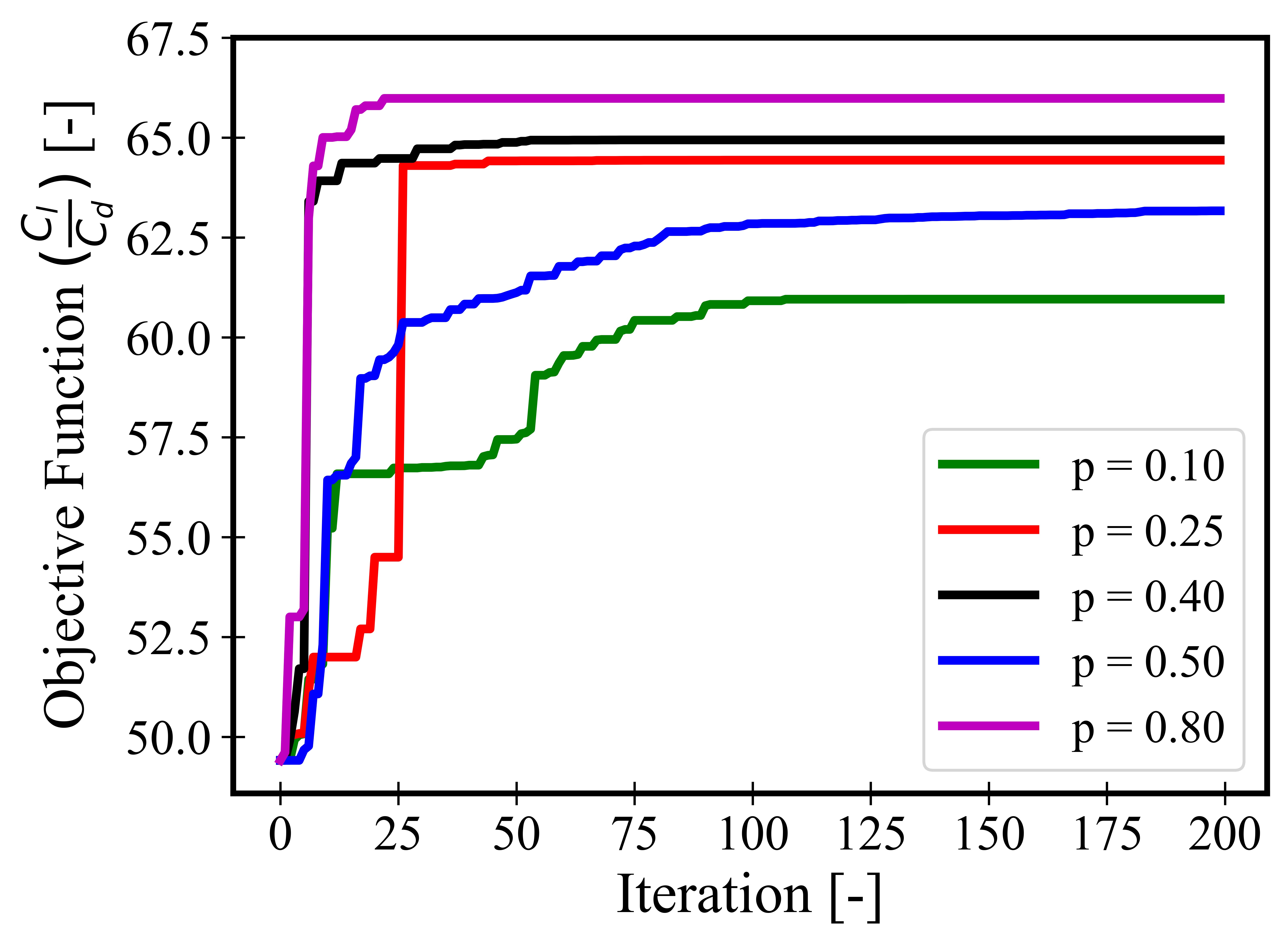} &   
  \includegraphics[width=65mm]{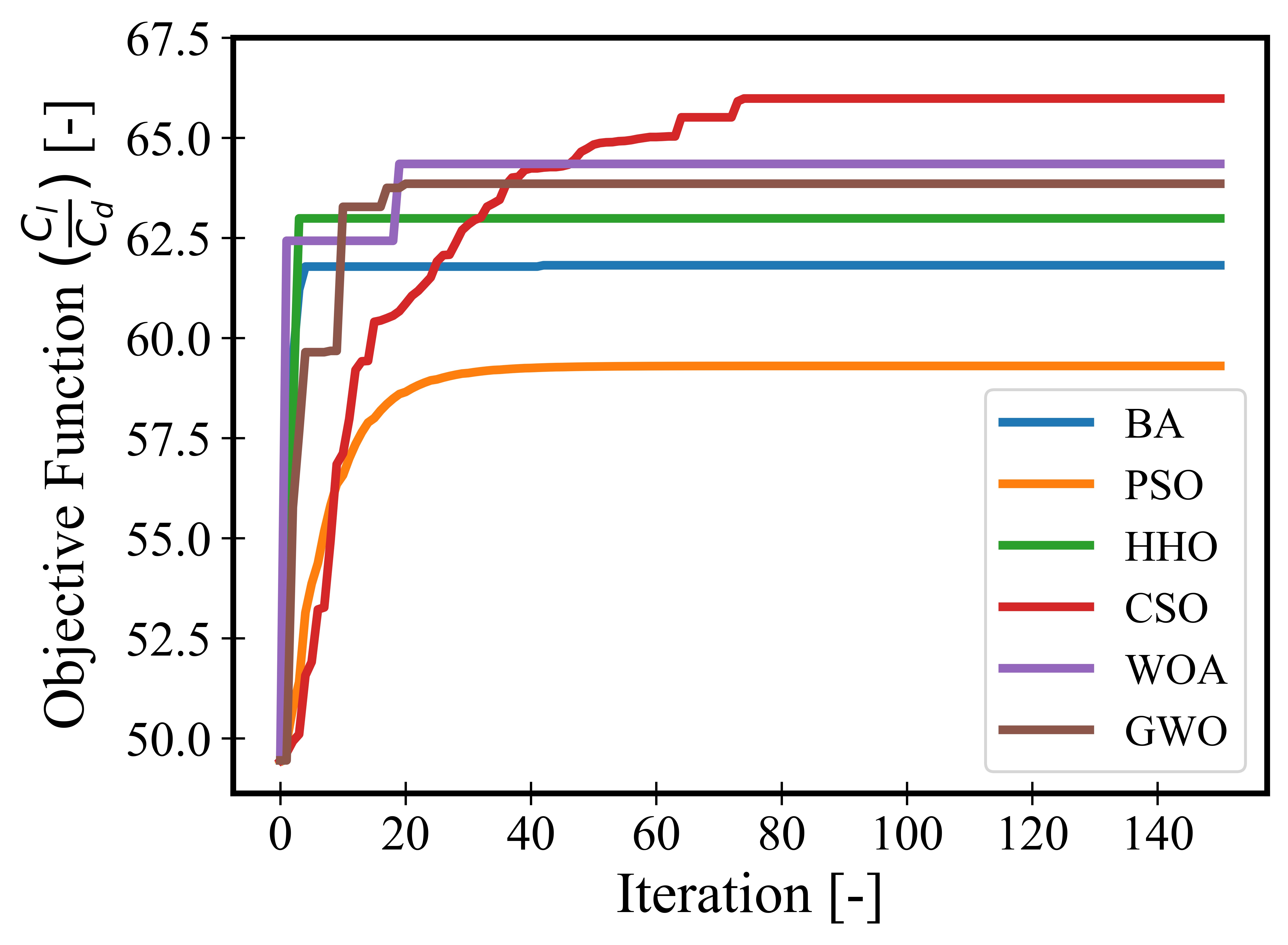} \\
(a) & (b) \\[6pt]
\end{tabular}
\caption{(a) Performance of CS algorithm for different values of probability parameter (b) Performance of other metaheuristic algorithms compared to CS algorithm}
\label{params}
\end{figure}

After the RBF surrogate model has been trained with the CFD data, Cuckoo Search optimization is performed on the response surface to maximize the objective function. As discussed earlier, the CS algorithm has a probability parameter that is responsible for the exploratory phase of the algorithm. The probability parameter must be chosen carefully after a thorough inspection to ensure that the algorithm performs most efficiently. 
Figure \ref{params}(a) illustrates the impact of various probability parameters on the algorithm's performance. As the algorithm operates stochastically, it was executed 20 times for each probability parameter, and the mean performance of each parameter was recorded and is shown in the figure. The convergence of the CS algorithm was compared for the following values of $p_a$, 0.1, 0.25, 0.4, 0.5, and 0.8. It can be observed in Figure \ref{params} that the CS algorithm performs best for $p_a$ = 0.8, as it converges much quicker than and also obtains the most optimal solution. At $p_a$ = 0.4, the algorithm converges at a similar rate but seems to move towards a local minimum and obtains a sub-optimal solution. For $p_a$ = 0.1, the optimal solution is obtained but at a much slower rate when compared to $p_a$ = 0.8. For $p_a$ = 0.25 and 0.5, the algorithm fails to obtain an optimal solution. Therefore, in this study, $p_a$ = 0.8 is observed to generate the best design for the gurney flap.

\subsection{Performance Against Other Algorithms}
The results obtained by the CS algorithm are quantitatively compared against other optimization methods to objectively assess its performance. For increased computational efficiency, $N_p$ was set to 5 for all algorithms. The following metaheuristic algorithms were used for comparison:

\begin{itemize}
    \item Particle Swarm Optimisation (PSO): PSO involves the swarm intelligence of bird flocks. Owing to its simplicity and efficiency, it has become a tool to solve optimization problems \citep{pso}. 
    \item Harris' Hawk Optimizations (HHO): HHO is a novel, complex meta-heuristic approach that utilizes the intelligence of hawks in attacking and chasing their prey \citep{hho}.
    \item Whale Optimization Algorithm (WOA): WOA is a novel swarm-based algorithm based on humpback whale's hunting behavior \citep{woa}. 
    \item Bat Algorithm (BA): BA is an algorithm that mimics the echolocation ability of micro bats for solving mathematical problems \citep{ba}.
    \item Grey Wolf Optimization (GWO): GWO is an optimization algorithm inspired by wolves' hunting behavior \citep{gwo}. 
\end{itemize}

As we can see in Figure \ref{params}(b), the highest percentage of improvement in the objective function is obtained by the Cuckoo Search algorithm, followed closely by WOA and GWO. The better results obtained by the CS algorithm can be attributed to the superior exploration capabilities of CS that help it generate new solutions to escape from local optima. CS has a good exploration capability due to its ability to generate new solutions using random walk steps and the Lévy flight step. In contrast, PSO and HHO are reliant on the current best-known solutions, making them prone to getting trapped in local optima. GWO’s premature convergence can be explained by considering the dominance of alpha and beta wolves. BA and WOA do not show as good an exploration capability since they rely mainly on the random search and adaptive movement, respectively.
Also, the CS algorithm has a mechanism to maintain diversity by replacing the worst solutions with new ones. This helps to prevent the search from getting stuck in a suboptimal region. PSO and HHO lack such a mechanism, which can limit their ability to explore the search space. GWO has a diversity mechanism, but it may be insufficient in some cases.

\subsection{Convergence Monitors}

\begin{figure}[h]
 
\begin{tabular}{cc}
  \includegraphics[width=65mm]{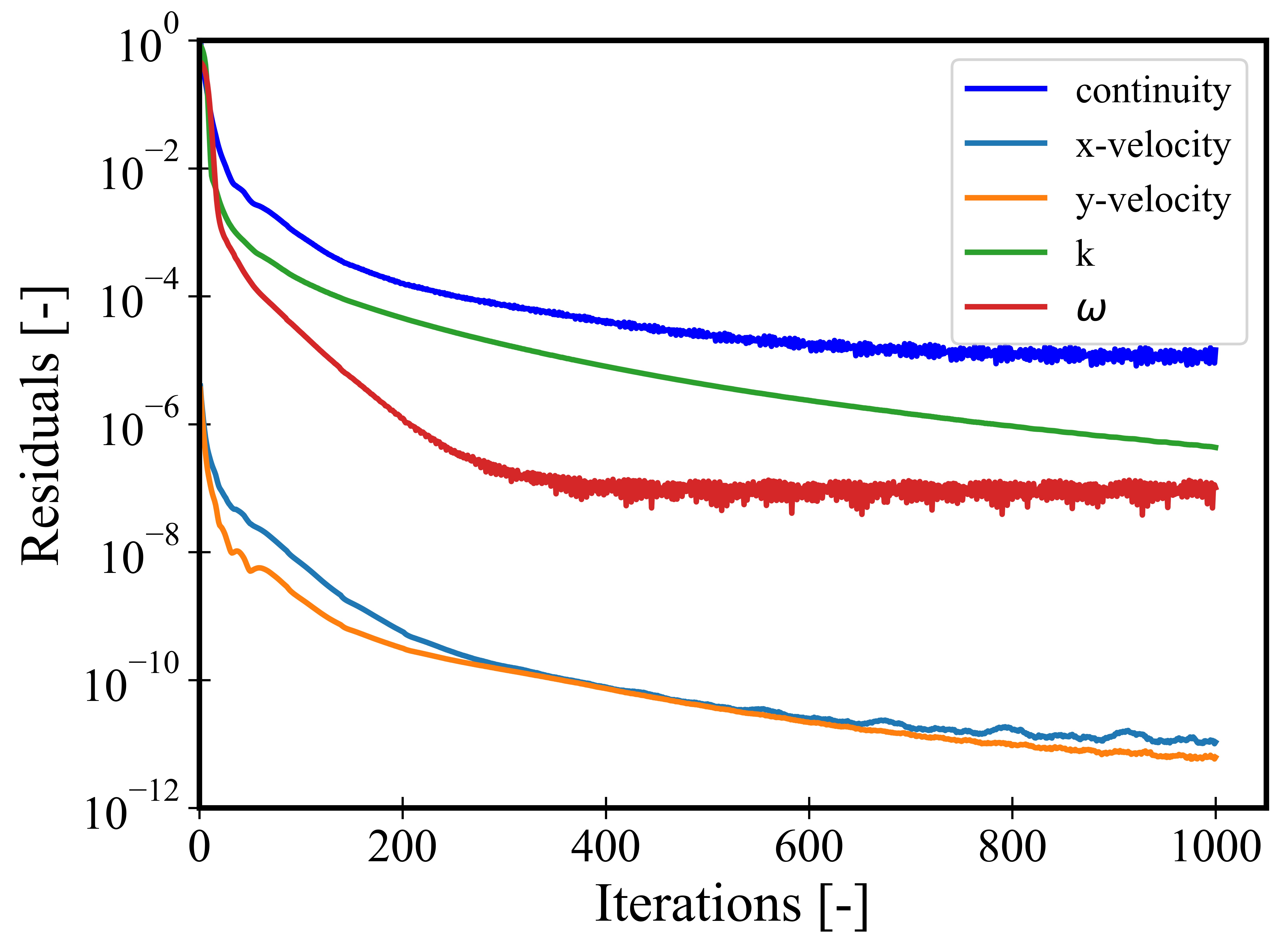} &   
  \includegraphics[width=65mm]{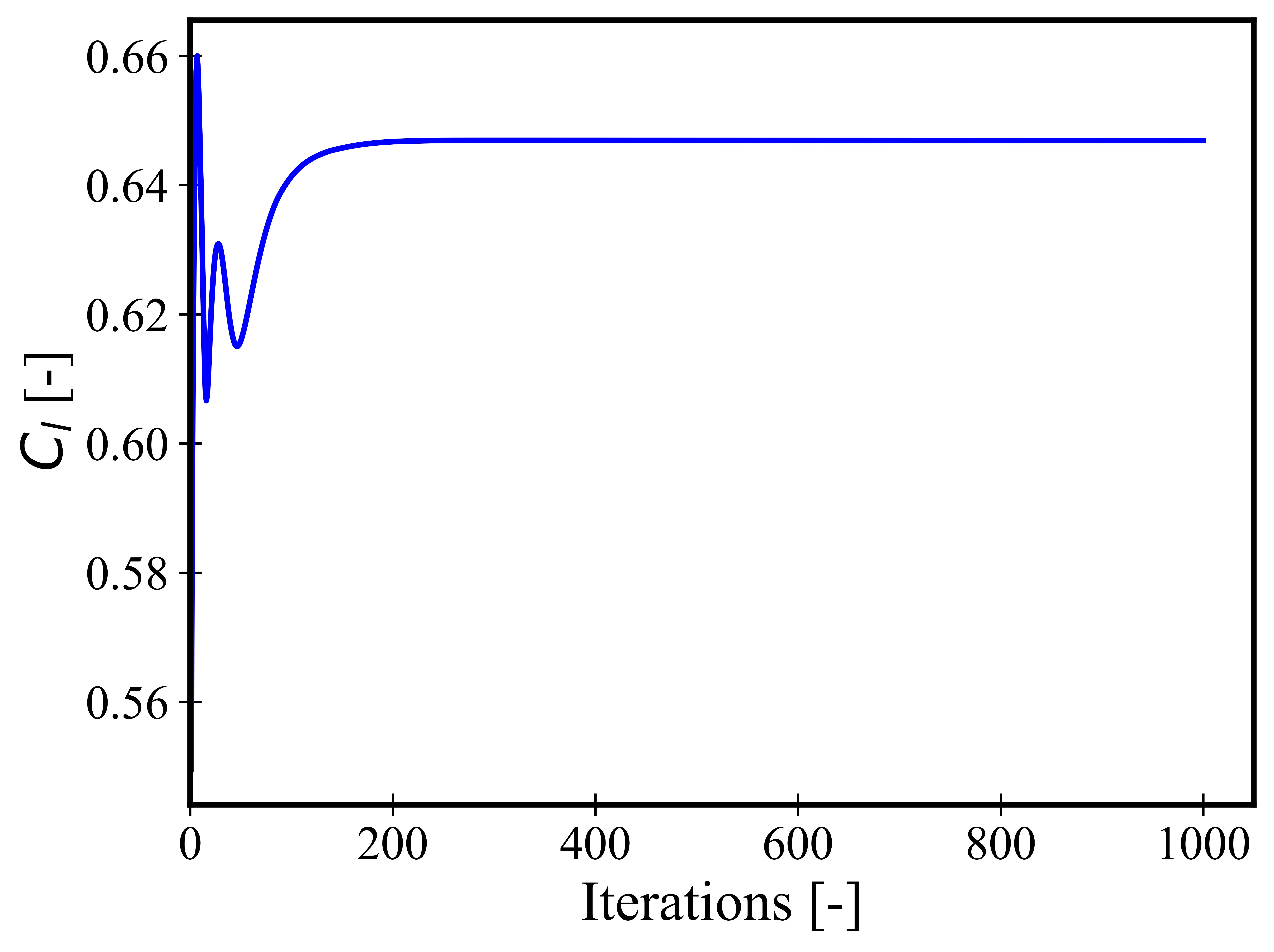} \\
(a) & (b) \\[6pt]
  \includegraphics[width=65mm]{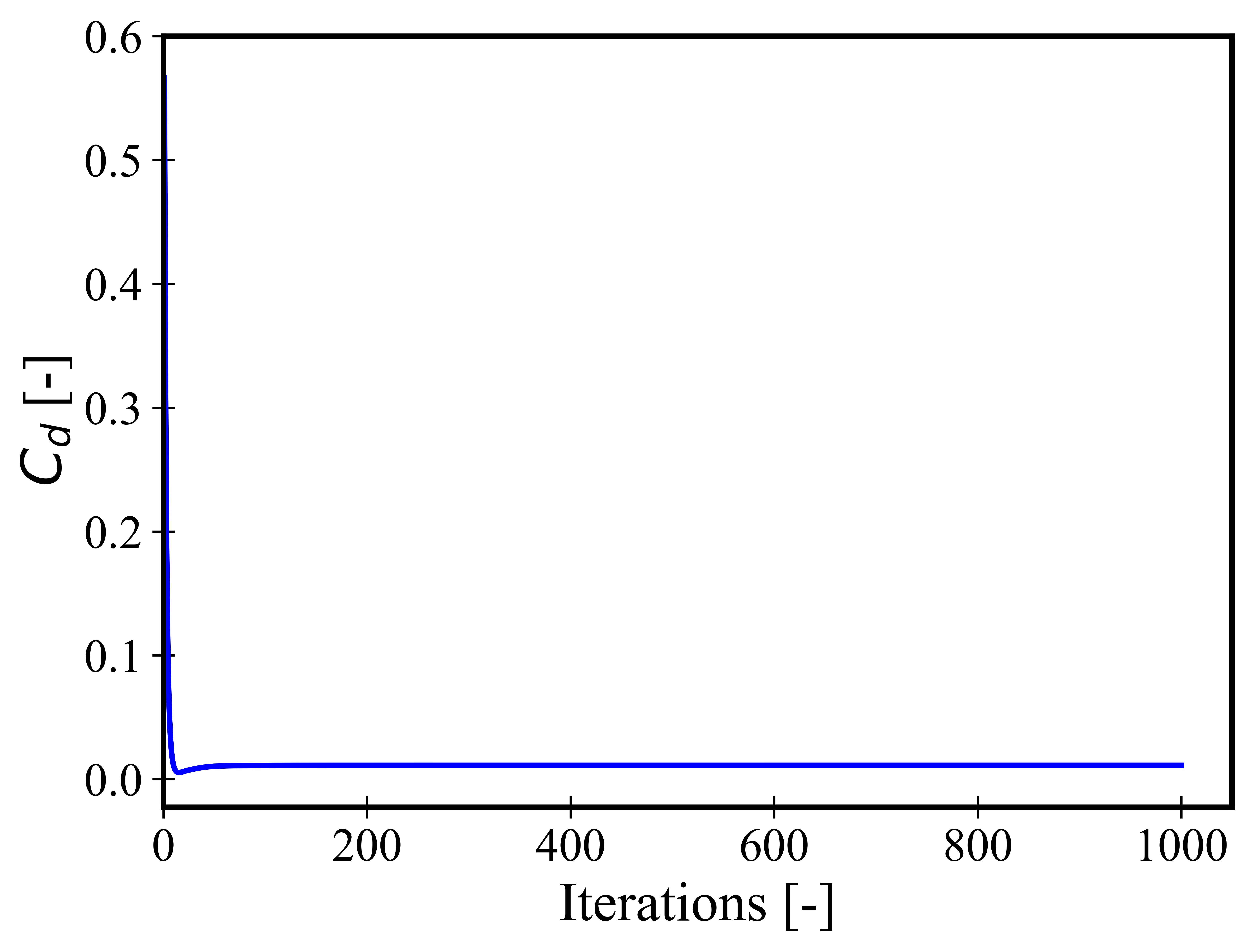} &
  \includegraphics[width=65mm]{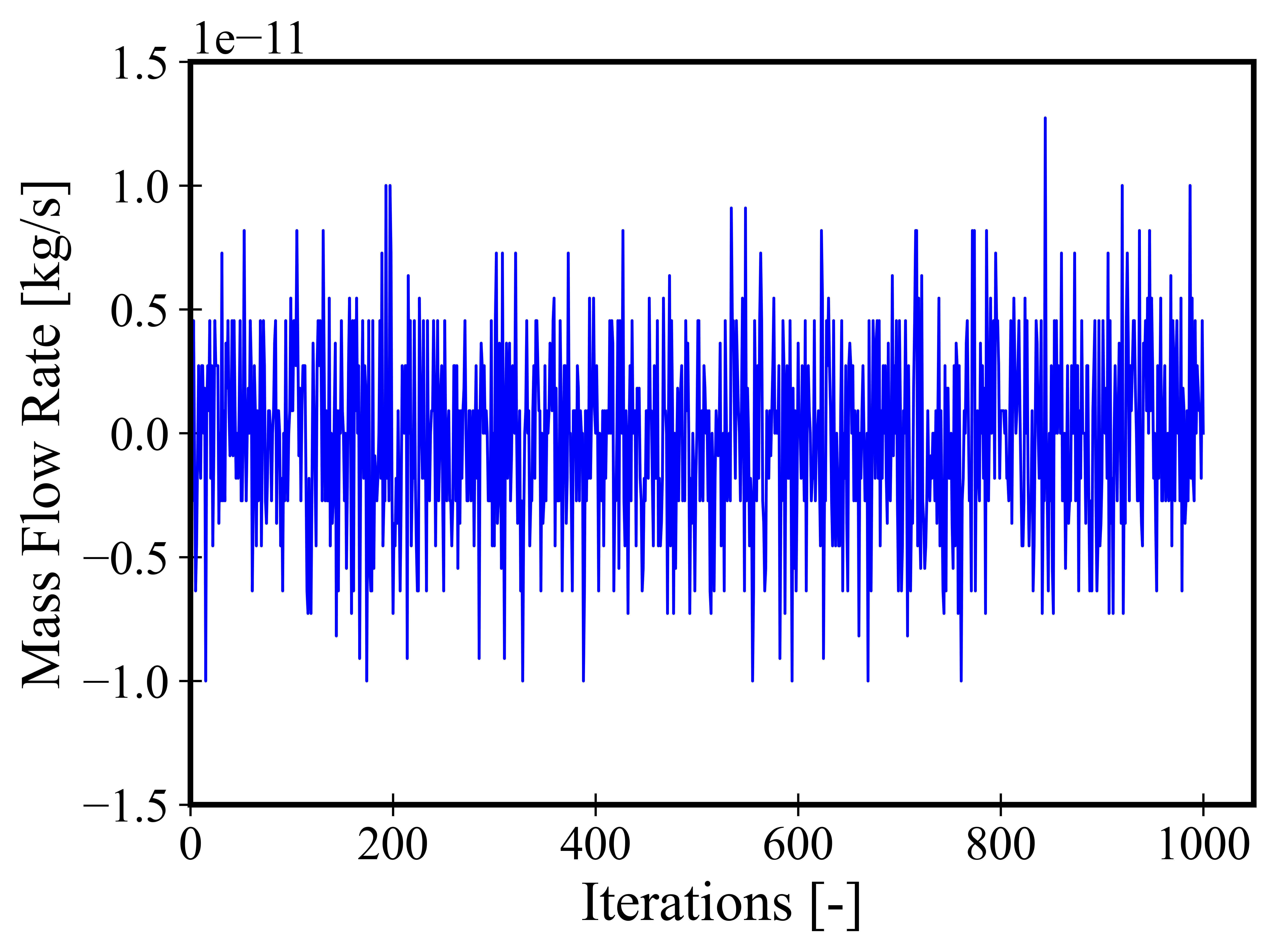}\\
(c) & (d) \\[6pt]
\end{tabular}
\caption{Convergence monitors for Foil A. (a) Residuals, (b) $C_l$, (c) $C_d$, and (d) Global mass imbalance}
\label{convergence}
\end{figure}

Convergence monitoring was employed during the steady-state simulations, where the iteration count was set at 1000, and a combination of convergence monitors was used. These included scaled residuals for continuity, momentum, k, $\omega$, computed values of global mass imbalance, and the coefficients of lift and drag ($C_l$ and $C_d$). The convergence monitors for the Foil A case are presented in Figure \ref{convergence}, indicating that the residuals fall below the tolerance limit of $1\times10^{-5}$ within the pre-set 1000 iterations. The constant values attained by $C_l$ and $C_d$ suggest that the simulation had converged to a stable solution. Moreover, the net outflow of mass settled at a value in close proximity to zero well before the completion of 1000 iterations. These results, in conjunction with the comprehensive grid dependence and validation studies conducted in section \ref{Sec2}, supports the accuracy of the numerical model in capturing the flow physics of the current study. These findings provide valuable insights for future studies seeking to predict and analyze the aerodynamic performance of Gurney flaps. 
\begin{table}[h]
    \centering
    \renewcommand{\arraystretch}{1.2}
    \begin{tabular}{c c c c c}
        \hline
        Case & $C_d$ $[-]$ & $C_l$ $[-]$ & $C_l/C_d$ $[-]$ & Change in $C_l/C_d$ [\%]\\
        \hline
        Foil A & 0.011 & 0.647 & 58.818 & 0  \\
        Foil B & 0.025 & 1.223 & 48.920 & -16.838 \\
        Foil C & 0.015 & 0.973 & 64.867 & +10.284\\
        \hline
    \end{tabular}
    \caption{Comparison of force coefficients between Foil A, Foil B, and Foil C}
    \label{resultsTable}
\end{table}
\subsection{Optimum Gurney Flap Parameters}
The findings of the optimization study are summarized in Table \ref{resultsTable}, where Foil A represents the baseline airfoil without the GF, Foil B represents the initial design point ($h=35$ mm and $\theta = 8 \degree$) of the optimization algorithm, and Foil C represents the optimum GF configuration (h=19 mm and $\theta = -58\degree$). It can be observed that the implementation of the GF in the Foil B configuration significantly increases the value of $C_l$ compared to Foil A, but at the same time, the value of $C_d$ also increases considerably, resulting in a decrement in the value of $C_l/C_d$ compared to Foil A. Conversely, for the Foil C configuration, while the value of $C_l$ increases significantly compared to Foil A, the increase in the value of $C_d$ is relatively low, leading to an increase in $C_l/C_d$ compared to Foil A. Overall, the results demonstrate that the optimized Gurney flap configuration (Foil C) offers a substantial improvement in airfoil efficiency (measured by $C_l/C_d$) compared to the baseline (Foil A) and the initial design point (Foil B).

\subsection{Trailing Edge Flow Structure}
\begin{figure}[h]
\begin{tabular}{cc}
  \includegraphics[width=65mm]{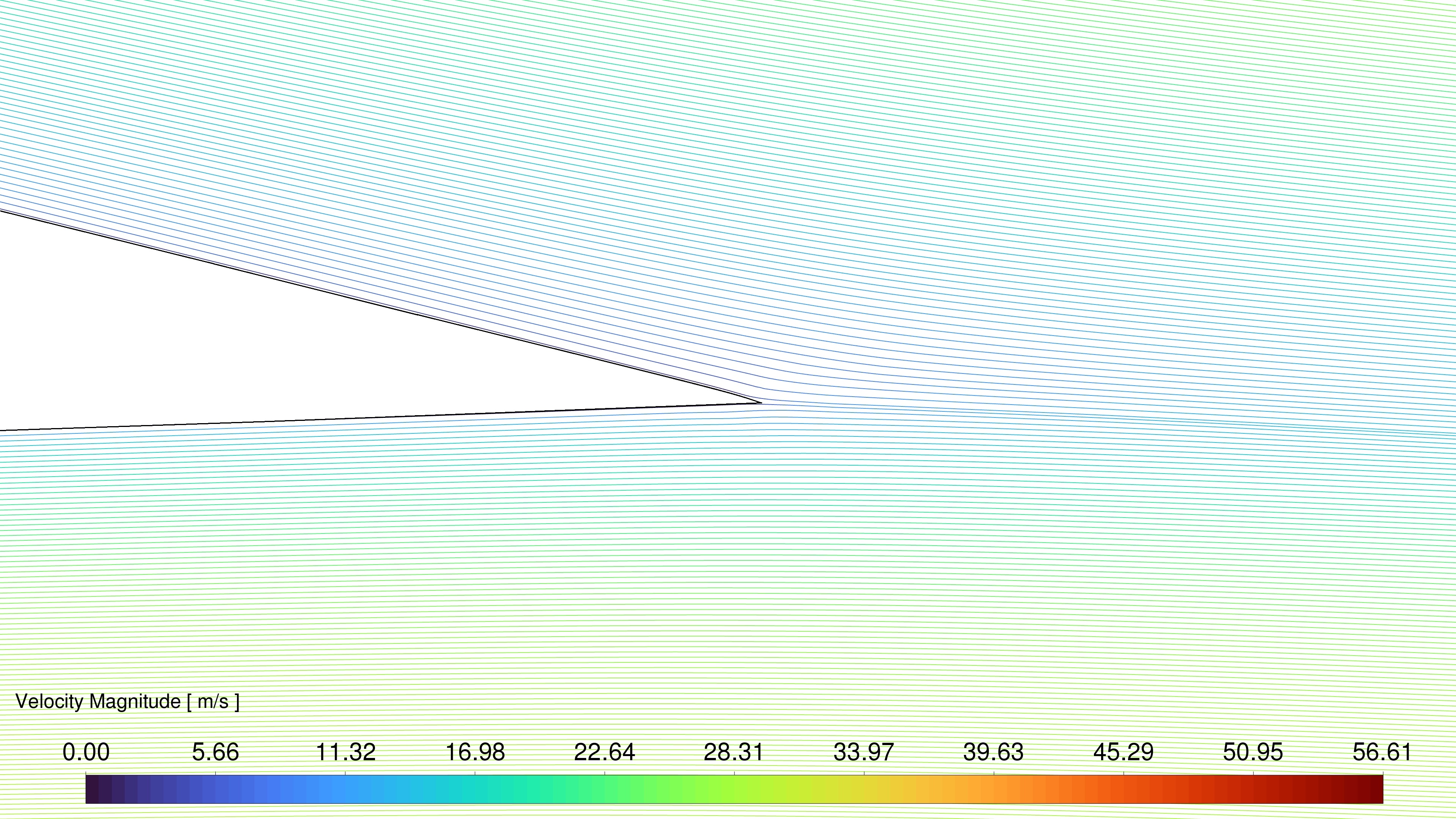} &   \includegraphics[width=65mm]{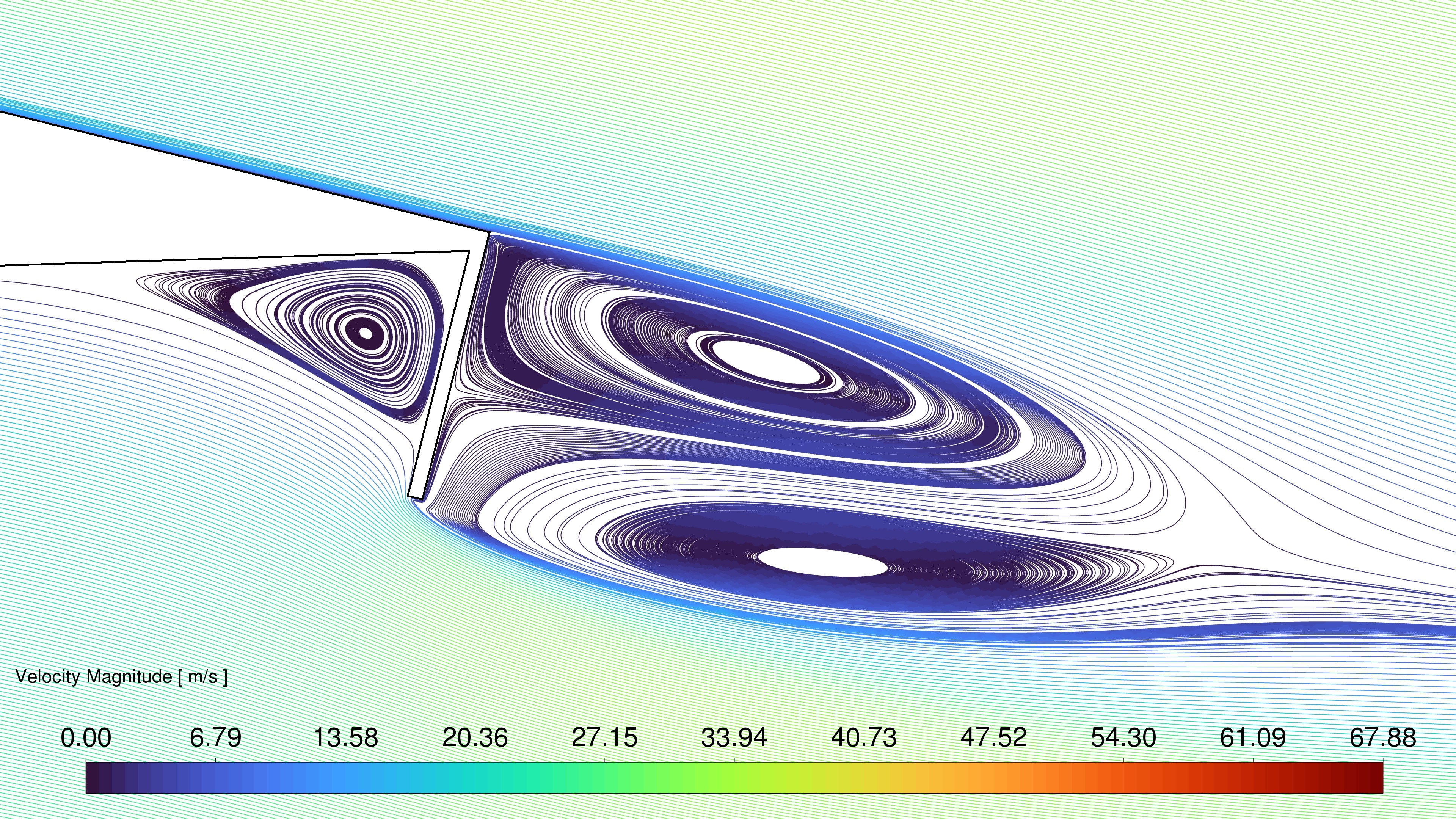} \\
(a) & (b) \\[6pt]
\multicolumn{2}{c}{\includegraphics[width=65mm]{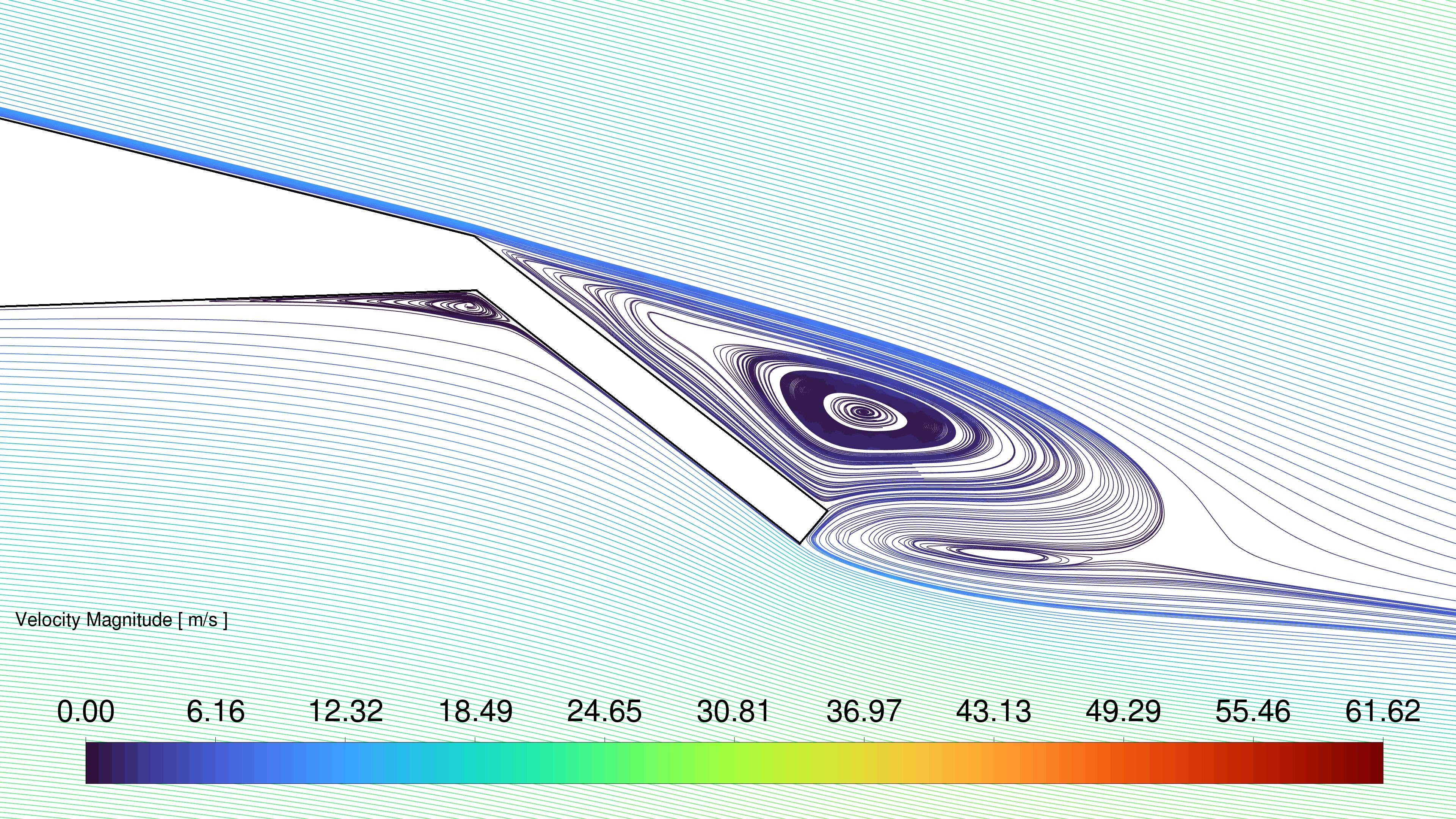} }\\
\multicolumn{2}{c}{(c)}
\end{tabular}
\caption{Computed pathlines in the vicinity of the trailing edge of NACA 0012 airfoil at AoA= $6\degree$ for (a) Foil A, (b) Foil B, and (c) Foil C}
\label{streamlines}
\end{figure}

The pathlines in the vicinity of the trailing edge (AoA=6$\degree$) are shown in Figure \ref{streamlines}. For Foil B and C, recirculation zones can be observed upstream and downstream of the GF consistent with Liebeck’s hypothesis \citep{Liebeck1978}. In addition, a downwash of air at the trailing edge is present due to the GF. The recirculation zone downstream of the GF is inclined downwards and shifts the stagnation point far aft of the flap, increasing the effective chord and camber of the airfoil. This ultimately suggests that instead of using a longer and more cambered airfoil, one can get the same performance with a smaller airfoil equipped with a GF. From Figures \ref{streamlines}(b) and \ref{streamlines}(c), it can be inferred that Foil B has a larger recirculation zone and a stronger down-washing effect compared to Foil C thus explaining its higher lift ($C_l$) and drag ($C_d$) coefficients and lower airfoil efficiency ($C_l/C_d$).

\subsection{Velocity Distribution}

\begin{figure}[p]
\centering
\begin{tabular}{cc}
  \includegraphics[width=115mm]{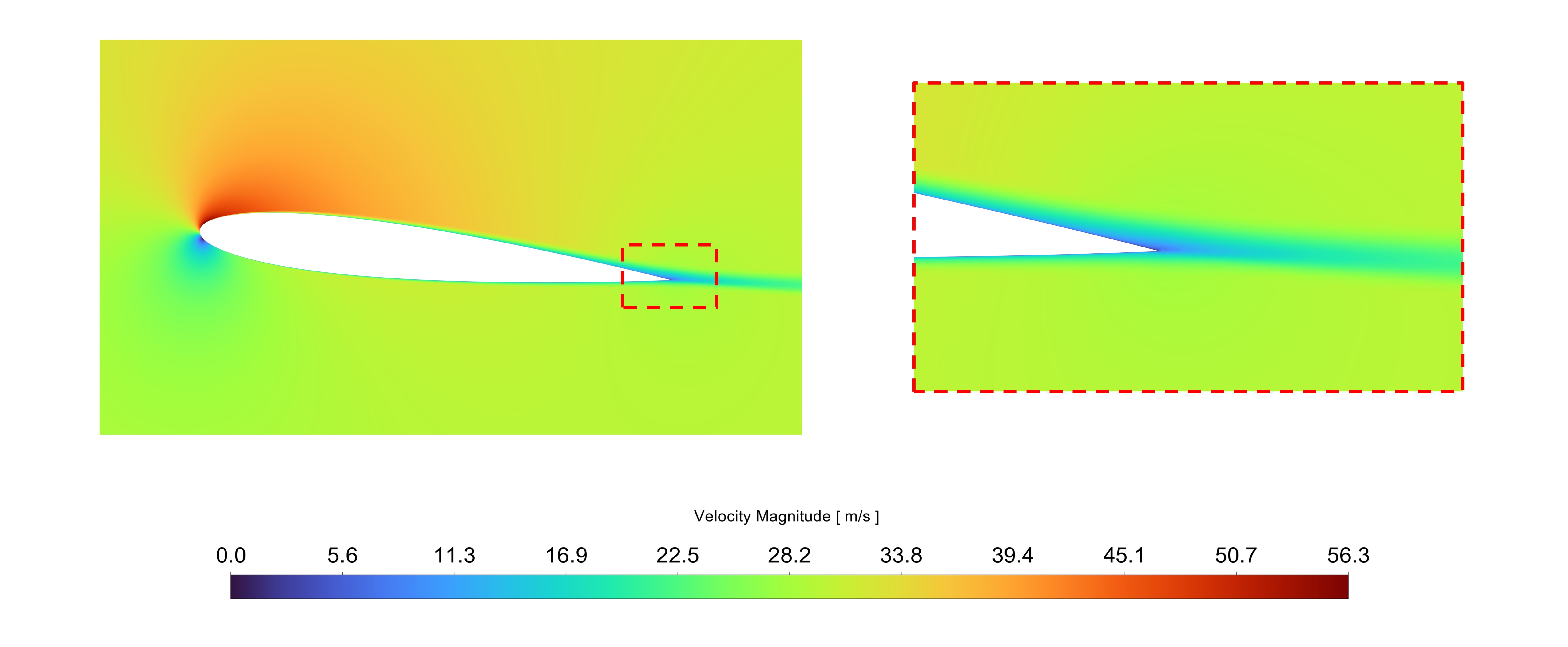} \\
(a) \\[6pt]
  \includegraphics[width=115mm]{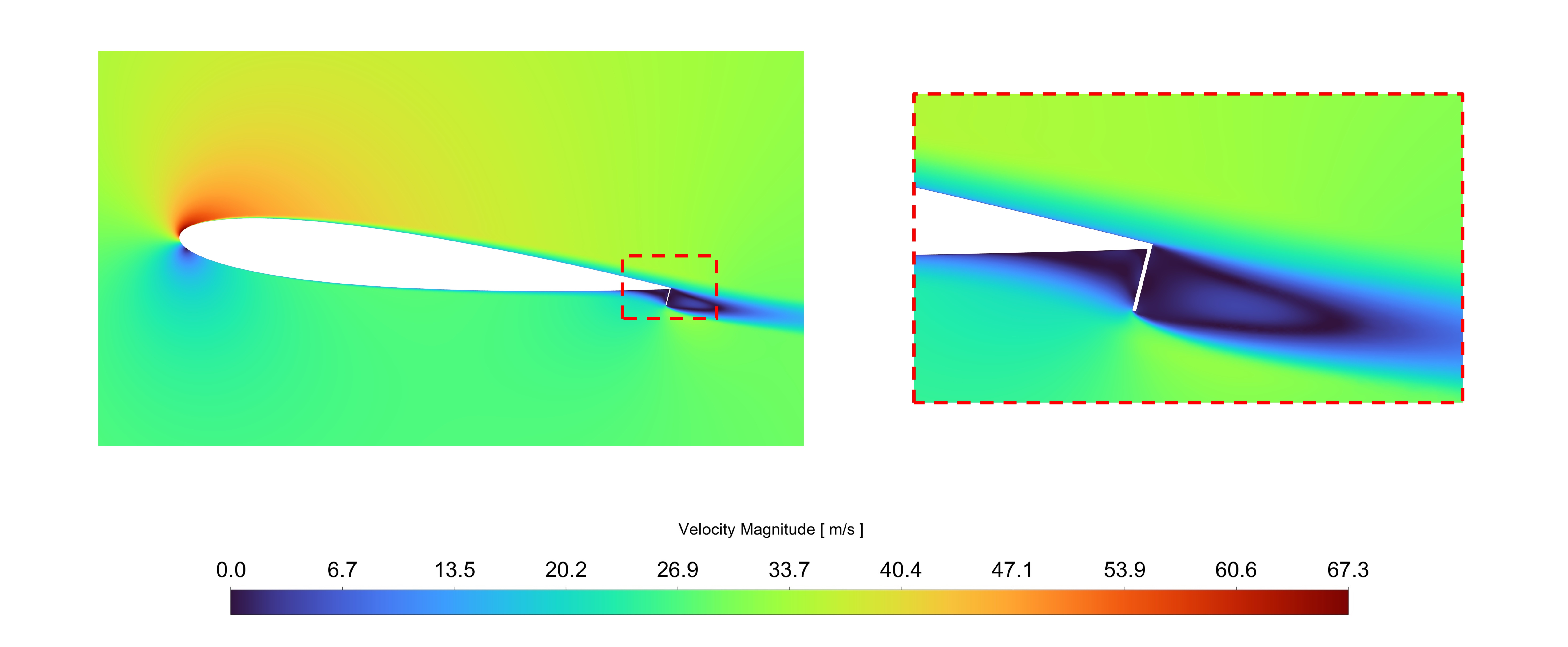}\\
(b) \\[6pt]
  \includegraphics[width=115mm]{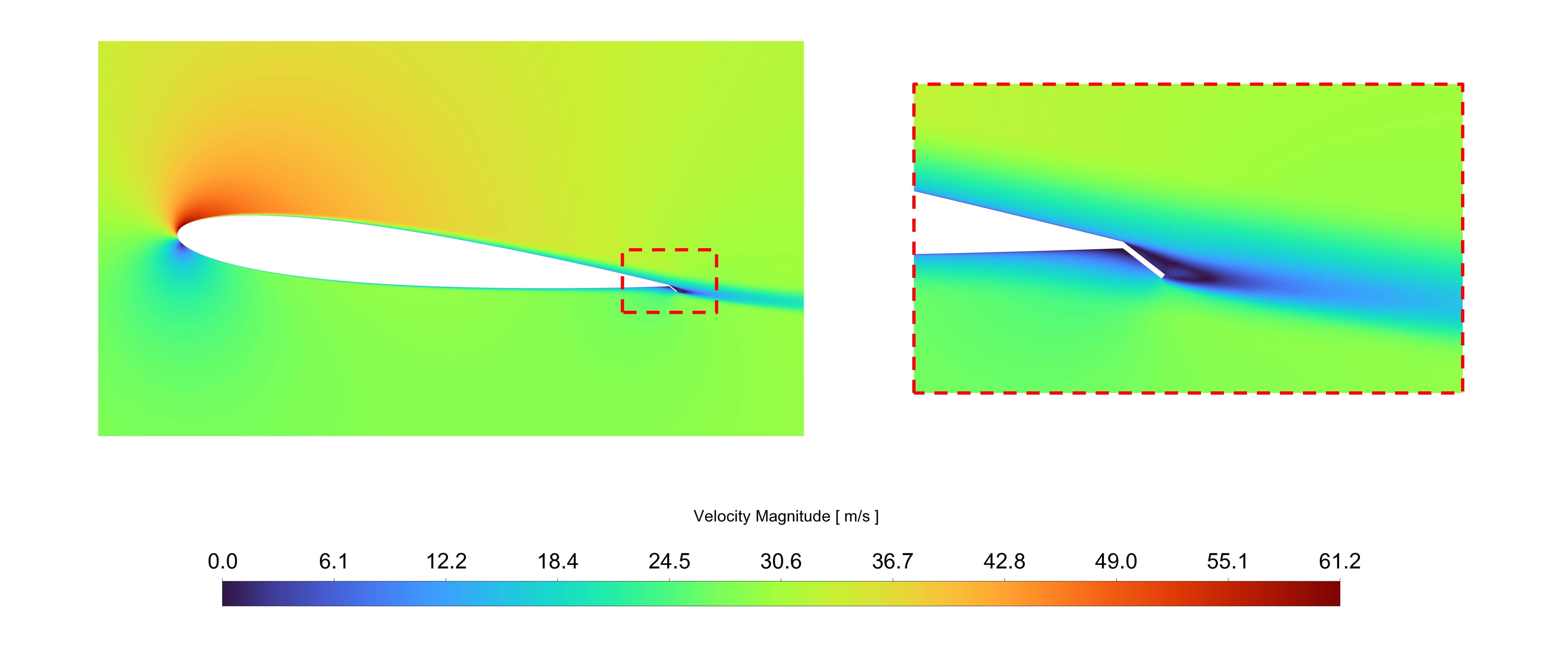}\\
(c) \\[6pt]
\end{tabular}
\caption{Velocity contours in the vicinity of NACA 0012 airfoil at AoA= $6\degree$ for (a) Foil A, (b) Foil B, and (c) Foil C}
\label{velocity}
\end{figure}
From the velocity contours shown in Figure \ref{velocity}, it can be observed that there is a greater momentum deficit in the wake of Foil B compared to Foil A and Foil C. With the use of a larger GF, the flow over the suction surface is accelerated to higher velocities, thus increasing the suction pressure. Also, in the case of Foil B, the steep mounting angle of the GF with the chord stagnates the incoming air beneath the airfoil near the windward side of the flap thus raising the pressure on the pressure side and ultimately contributing towards an increment in the lift force that the airfoil generates. But at the same time, the increased pressure on the windward side and decreased pressure on the leeward side of the GF increases the net pressure drag acting on Foil B compared to Foil A and Foil C and thus contributes to the increment in its drag coefficient ($C_d$) and a reduction in airfoil efficiency ($C_l/C_d$).

\subsection{Pressure Distribution}
\begin{figure}[p]
\centering
\begin{tabular}{cc}
  \includegraphics[width=115mm]{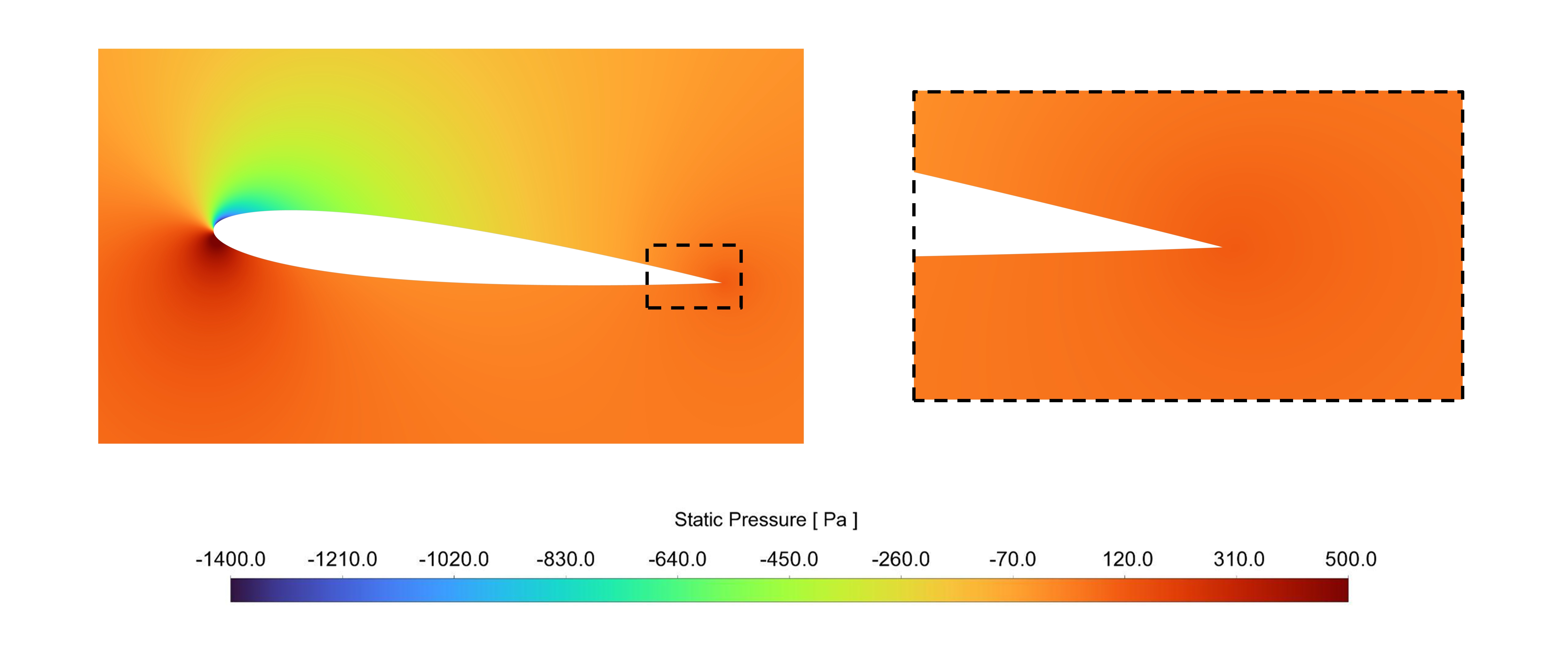} \\
(a) \\[6pt]
  \includegraphics[width=115mm]{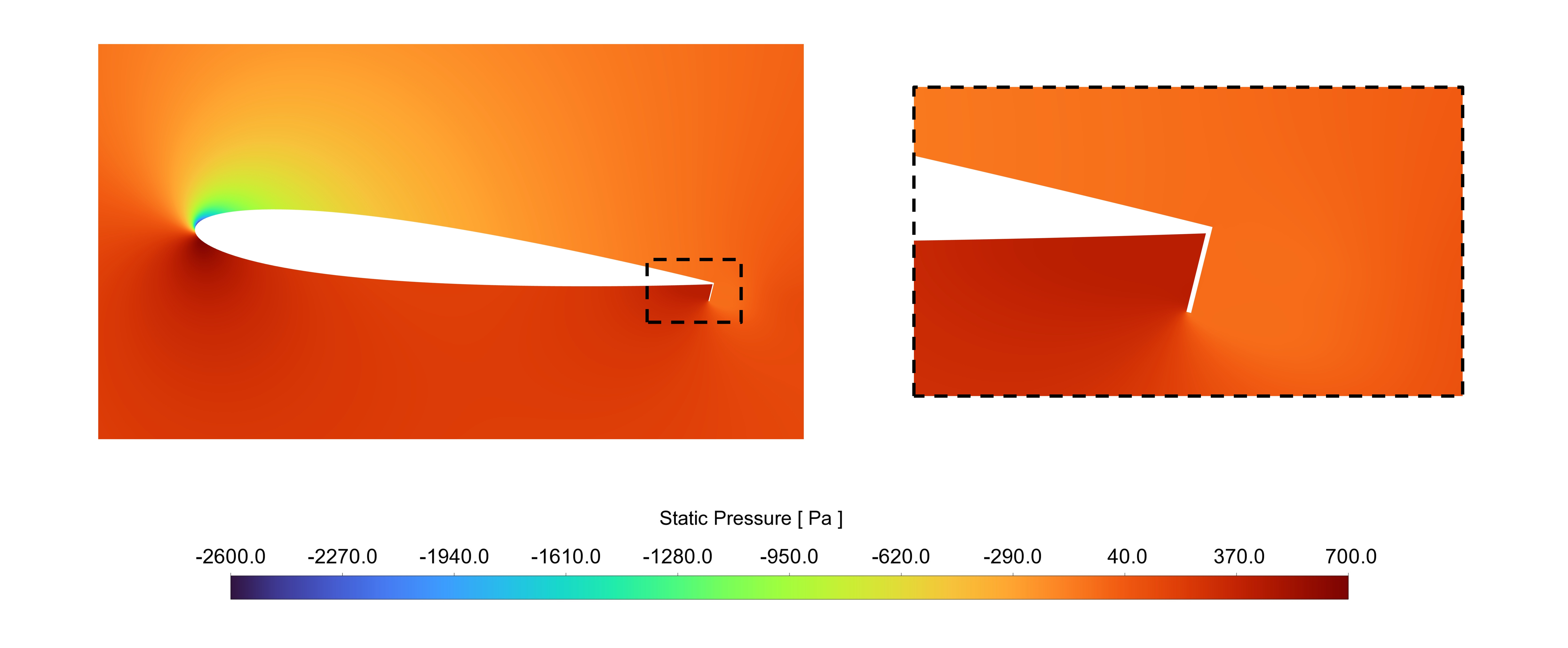}\\
(b) \\[6pt]
  \includegraphics[width=115mm]{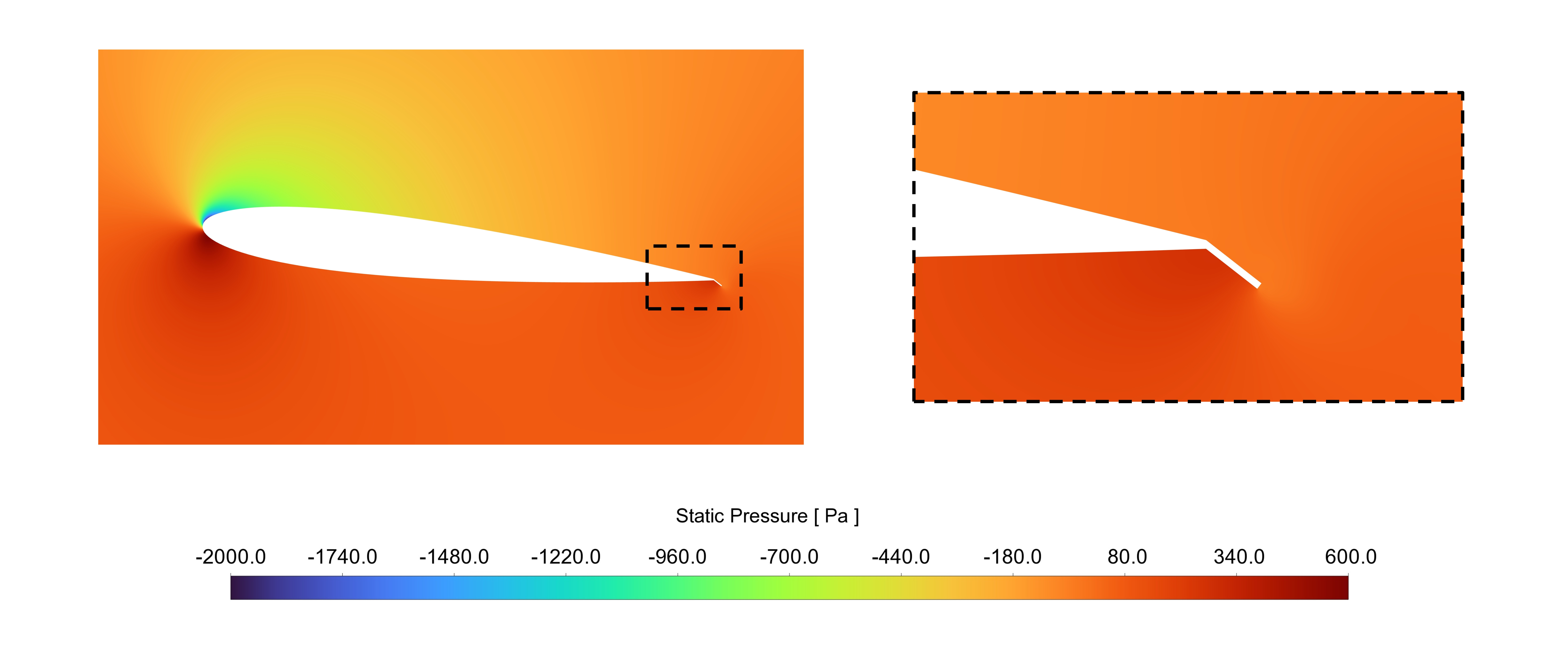}\\
(c) \\[6pt]
\end{tabular}
\caption{Pressure contours in the vicinity of NACA 0012 airfoil at AoA= $6\degree$ for (a) Foil A, (b) Foil B, and (c) Foil C}
\label{pressure}
\end{figure}
\vspace{2pt}
The inclusion of GF increases the pressure levels adjacent to the lower surface while concurrently reducing the pressure along the upper surface. This can be observed from the pressure contours shown in Figure \ref{pressure}.
The recirculation zone downstream of the GF (Figure \ref{streamlines}) generates a low-pressure region near the leeward side. This effectively mitigates the adverse pressure gradient near the rear end of the airfoil. As a result, flow separation over the suction surface of the airfoil at higher angles of attack is delayed or even eliminated. Also, for small flap heights ($h\leq 2\%c$), this effect reduces the stall angle of attack while also making the loss of lift near the stall region more gradual.
For Foil B, when the incoming airfoil beneath the airfoil encounters the flap, it stagnates due to the sharp change in its flow direction leading to larger increments in pressure at the bottom of the airfoil close to the windward side of GF. On the other hand, it is comparatively easier for the incoming air to follow the contour of the airfoil and the GF in the case of Foil C as its inclination is greater towards the leeward side, hence the pressure increase is less compared to Foil B, explaining its lower lift coefficient ($C_l$). 

This observation is further supported by the distribution of the pressure coefficient ($C_p$) along the airfoil, as depicted in Figure \ref{Cp_wake}(a). Clearly, along the suction surface, particularly near the leading and trailing edge, the suction pressure is more for Foil B followed by Foil C and Foil A. In addition, the highest pressure near the windward side of the GF is present in Foil B followed by Foil C and Foil A.

\subsection{Wake Velocity}
\begin{figure}[p]
 
\begin{tabular}{cc}
  \includegraphics[width=65mm]{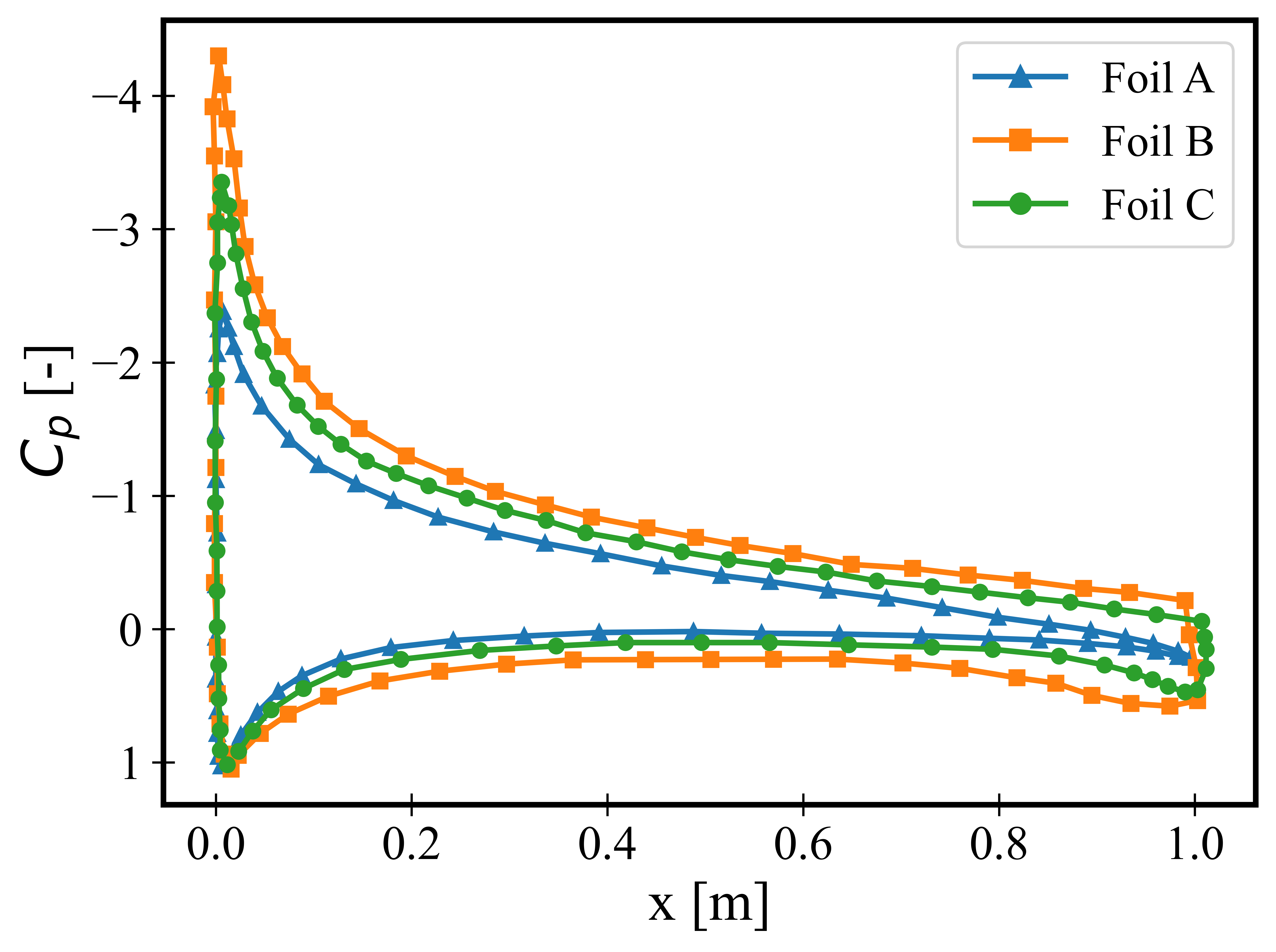} &   
  \includegraphics[width=65mm]{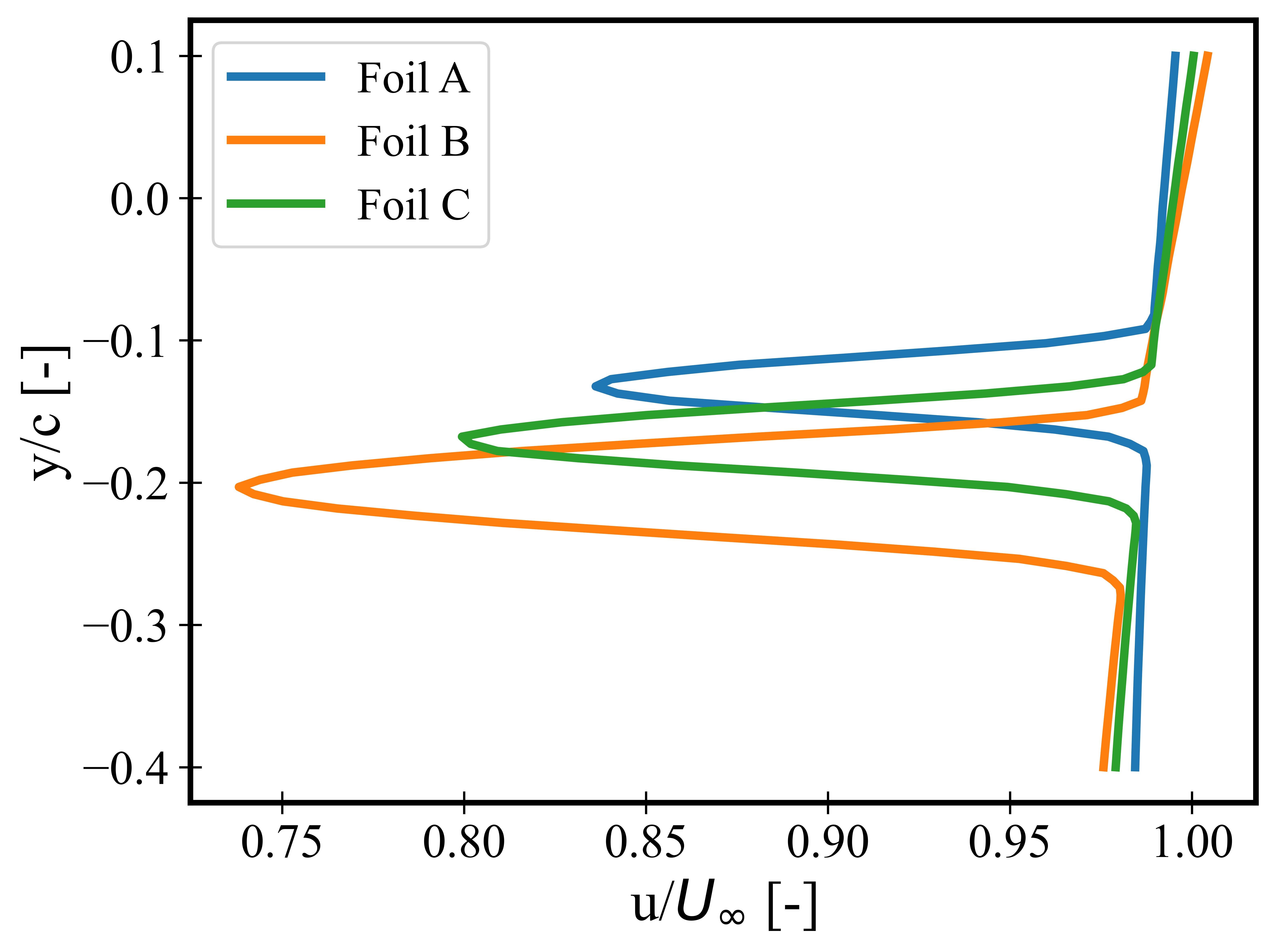} \\
(a) & (b) \\[6pt]
\end{tabular}
\caption{(a) Pressure distribution along the surface of NACA 0012 airfoil (AoA= $6\degree$) for Foil A, B, and C and (b) Wake velocity profiles at x=1.7\%c downstream of the leading edge of NACA 0012 airfoil (AoA= $6\degree$) for Foil A, B, and C }
\label{Cp_wake}
\end{figure}

The wake velocity distribution at a distance of 1.7c from the airfoil's leading edge for Foil A, B, and C at AoA=$6\degree$ is shown in Figure \ref{Cp_wake}(b). The highest momentum deficit in the wake region is observed for Foil B, followed by Foil C and Foil A, respectively. This effect is due to the stronger recirculation zone in the case of Flap B which extracts more energy from the incoming flow. By reducing the flap height and increasing its inclination from the vertical towards the leeward side, the size of the recirculation region is reduced, thus reducing the dissipation of energy in that area. This results in a greater wake velocity in the case of Foil C compared to Foil B. There is an absence of any recirculation zone at the trailing edge of Foil A which explains the presence of greater momentum in its wake.
Also, it can be observed that there is a shift in the wake profiles in the downward direction for airfoils equipped with GF. This shift is greater for Foil B compared to Foil C indicating that Foil B shifts the Kutta condition further downwards. The bending of flow due to the presence of GF is equivalent to the increment in the airfoil's trailing edge camber, as documented by \citet{Neuhart1988AWT}. Hence Foil B has a greater effective camber in comparison to Foil C leading to a larger lift coefficient ($C_l$). But due to the larger momentum deficit in the wake, the drag coefficient ($C_d$) of Foil B is greater in comparison to Foil C. 

\section{Conclusion}
This study proposed an optimization framework for a GF mounted on a NACA 0012 airfoil. The use of RBF neural network as a surrogate model in the framework resulted in a notable enhancement in computational efficiency. The Cuckoo Search algorithm was found to be superior to other state-of-the-art algorithms for attaining optimal design parameters for the Gurney flap. The shift in the trailing edge Kutta condition vertically downwards and away from the flap towards the leeward side for airfoils equipped with Gurney flaps suggests an increment in the effective camber and chord. The optimal Gurney flap configuration was found to have a flap height of 1.9\%c and a flap mounting angle of $-58\degree$, resulting in an improvement in the airfoil's  efficiency ($C_l/C_d$) by 10.28\% compared to the baseline airfoil.
However, there were some limitations to this study. The use of only a section of a wing ignored the flow change effects in the spanwise direction. Additionally, the optimization framework only considered parameters for flap height and mounting angle, without exploring changes in the overall shape of the Gurney flaps, which limits the potential for greater performance gains. The use of steady-state RANS simulations, due to limited computational resources, also limits the accuracy of the results compared to high-fidelity LES and DNS simulations. The use of surrogate models can also lead to less accurate results and in future studies, improved optimization algorithms can be used to obtain better results.\\
In addition to its application in the optimization of the Gurney Flap, the proposed framework exhibits potential for broader utilization in the design optimization of various other aerodynamic elements employed in automotive, aerospace, and wind engineering applications. Considering the underlying principles and methodologies employed within the proposed framework, it can be assumed that the framework's applicability extends beyond the immediate scope of the current research. 

\clearpage
\bibliographystyle{elsarticle-num-names} 
\bibliography{cas-refs}





\end{document}